# How Well Do LLMs Simulate Survey Responses Following a Breast Cancer Screening Intervention?

Kenneth Koh [1] kenneth_koh@mymail.sutd.edu.sg
Ryan Jak Yang Lim [2] ryan.limjy@u.nus.edu
Alessandro Sparacio [3] Alessandro_Sparacio@a-star.edu.sg
Peh Joo Ho [4,5] ephhpj@nus.edu.sg
Mile Sikic [1] miles@a-star.edu.sg
Borame L Dickens [4] ephdbsl@nus.edu.sg
Mikael Hartman [4,5,6] ephbamh@nus.edu.sg
Jingmei Li [1,5,7] lijm1@a-star.edu.sg

1 Genome Institute of Singapore (GIS), Agency for Science, Technology and Research (A*STAR), 60 Biopolis St, Genome, #02-01, Singapore 138672, Singapore

2 Department of Statistics and Data Science, National University of Singapore, 6 Science Drive 2, Singapore, 117546, Singapore

3 Institute for Human Development and Potential (IHDP), Agency for Science, Technology and Research (A*STAR), Brenner Centre for Molecular Medicine, 30 Medical Drive, Singapore 117609, Singapore

4 Saw Swee Hock School of Public Health, National University of Singapore, Tahir Foundation Building, 12 Science Drive 2, Singapore 117549, Singapore

5 Department of Surgery, Yong Loo Lin School of Medicine, National University of Singapore and National University Health System, Singapore, Singapore

6 Department of Surgery, National University Hospital and National University Health System, Singapore 119228, Singapore

7 National Cancer Centre Singapore (NCCS), Singapore Health Services (SingHealth), Singapore, Singapore

✉ Correspondence to: Jingmei Li Genome Institute of Singapore (GIS), Agency for Science, Technology and Research (A*STAR), 60 Biopolis St, Genome, #02-01, Singapore 138672, Singapore lijm1@a-star.edu.sg

# Abstract

**Background**
Collecting survey data is laborious and limited by privacy constraints. Large language models (LLMs) have shown promise as predictive social simulations. It is unclear whether they can replicate population-level response distributions before and after a healthcare intervention.

**Objectives**
(1) To evaluate whether agents informed solely by pre-intervention profile information can simulate post-intervention responses
(2) To assess predictive performance across question themes and demographic subgroups
(3) To assess the sensitivity of results to model input parameters

**Methods**
Groups of LLM agents (n=50, chosen by resampling without replacement) were created using information derived from 4125 women aged 35-59 years. Agent conditions ranged from random predictions, zero-shot (without task-specific training), to agent profiles enriched with aggregate (derived from the full dataset) or individual-level (random 50) demographic characteristics and pre-intervention questionnaire responses. Predicted responses were then generated using Gemma 4 E4B and Qwen3.5 9B. We compared distributions of predicted and observed questionnaire responses using Total Variation Distance (TVD) and Normalized Wasserstein Distance (NWD) (overall and subsets by demographics and question themes). Linear mixed-effects models were then used to assess subgroup differences in prediction error. We also examined whether model input parameters affect prediction accuracy by using the same simulated population across paired runs, only changing the parameter of interest.

**Results**
Across both LLMs, profile-based agents improved distributional accuracy. Post-Intervention, Aggregate and Individual Profile agents achieved NWDs of 0.093/0.092 for Gemma and 0.118/0.110 for Qwen, compared with 0.230 for random predictions. Zero-shot NWDs were 0.222 (Gemma) and 0.181 (Qwen), while sampling n=50 participants yielded 0.037. Prediction errors were higher among participants aged 55–59 years (vs 35-39 years; β=0.02, 95% CI [0.01, 0.03]) and those living in private property (vs HDB 1-4 room flats; β=0.02, 95% CI [0.01, 0.03]). Prediction accuracy varied by question theme and LLM model, with the highest errors observed for cancer fatalism (mean TVD=0.367 for Gemma; 0.187 for Qwen) and post-intervention attitudes toward genetics (mean TVD=0.231 for Gemma; 0.357 for Qwen). Sensitivity analyses showed that performance was influenced by prompt template changes (mean ΔNWD: -0.007 to 0.008; -4.0% to 7.8%) and temperature hyperparameter (0.6-0.0, mean ΔNWD: -0.004 to 0.014; -3.3% to 10.5%), though profile agents still outperformed zero-shot baselines.

**Conclusions**
Our results show the potential of LLM-based agents to model behavioral responses to interventions *in silico*. Providing contextual information of participants improved predictions over zero-shot prompting. However, profiles containing additional information beyond demographics did not consistently outperform simpler ones (relative performance varied across models and input parameters). Certain cultural constructs and population groups also remain inadequately represented by the LLM models evaluated. Future work may include building behaviorally grounded and locally validated virtual populations.

**Keywords**
Large Language Models; LLM; AI; Breast Cancer Screening; Simulation; Survey; Policy Intervention; Synthetic Data; Singapore

# Introduction

Despite national screening programmes and subsidies under Healthier SG, breast cancer screening uptake amongst women in Singapore has remained low at 35.2% [1]. Existing methods such as regression analysis [2], surveys [3], and focus group discussions [4, 5] can identify sub-populations who are less likely to screen. However, these methods are primarily descriptive and have limited ability to predict screening behaviour under certain interventions.

The use of LLMs for social simulations has received increasing attention in recent years. Park et al. [6] found that generative LLM agents given individuals' two-hour interview transcripts could predict 1052 participants' survey responses with up to 86% accuracy when benchmarked against test-retest consistency. Also, Ashokkumar et al. [7] found that GPT-4 predictions of 469 treatment effects across 70 social science surveys were strongly correlated with actual outcomes (r=0.85). However, some key concerns remain regarding the usability of such simulations. Wang et al. [8] argue that LLMs flatten within-group heterogeneity due to their training process, where the cross-entropy function rewards LLMs for less heterogeneous outputs. Similarly, Anthis et al. [9] further identify challenges relating to diversity, bias, sycophancy, alienness, and generalization that must be addressed before LLM survey agents can be used to reliably model human behavior.

Thus far, existing research on LLMs for such use cases has focused mainly on North American contexts. Evidence from other geographical contexts suggests that LLMs face difficulties in capturing population heterogeneity [10, 11]. Existing papers with positive results were often found to underreport or omit these demographic subcategories [12]. Since breast cancer epidemiology and screening patterns vary across geographic and socio-economic contexts [2, 13] we assess whether LLM-based simulations can reproduce survey responses from a Singapore-based personalized risk-based breast cancer screening study [14]. To evaluate their potential for predicting responses in future settings, we further assess the *in silico* predictive performance across question themes and demographic subgroups, as well as the sensitivity of results to model input parameters.

# Methods

### 2.1 Simulation setup

In this study, we create a simulation framework (**Figure 1**) using LLM agents with profiles built from demographics and responses pre-intervention (i.e. receipt of personalized breast cancer risk report (BCRR)). Demographic data was derived from the BREAst screening Tailored for HEr (BREATHE) cohort of 4129 women aged 35-59 years old (median 48, IQR: 42-53) [14]. Four participants with null responses were excluded, while 17 participants who developed breast cancer within six months (excluded in the original study) were retained. These participants were retained as their profiles were still useful for our study objective of modelling participant responses. Aggregate frequency distributions were then generated for each demographic variable.

#### Generation of agents using aggregated information

Each LLM agent uses a profile independently drawn from the aggregate distribution for a given demographic variable (**Multimedia Appendix 1, Tables S1-S4**). This method preserves the marginal distributions of each variable but not the correlations [15]. Other methods, i.e. Iterative Proportional Fitting (IPF) exist but have little benefits since a starting seed matrix preserving associations between demographic variables is unavailable.

#### Generation of agents using profiles of individuals

As an alternative, we also create agents using individual profiles (i.e. demographic profiles sampled directly from the original cohort without replacement) to preserve correlations between demographic variables. For individual profiles, we keep the same variables as the aggregate profile (**Multimedia Appendix 1, Tables S3-S4**). This allows us to see if preserving correlations of variables (reflecting actual individuals) provides benefits over marginal distributions alone, which could result in an unrealistic combination of traits.

#### Comparators for prediction

Aggregate- and individual- profile agents were compared with: (1) a uniform random baseline, where each possible answer is selected with equal probability; (2) a zero-shot condition, in which the LLM receives no information other than the question itself; (3) an age-only condition in which only participant age was provided (BCRR is age-dependent due to different guidelines for different age groups); and (4) minimal profile consisting of age, ethnicity and housing type. We additionally included (5) a majority-response baseline, which uses the most frequent response for each question in the observed dataset; (6) variation from sampling n=50, which estimated error when using 50 observed participant responses to approximate full cohort; and (7) a 50/50 split-sample empirical marginal baseline, where question specific response distributions were estimated from a random half of the cohort and evaluated against the held-out half. (**Table 1**)

#### Information given to agent profiles

Pre-intervention agents were conditioned only on information available before baseline survey completion. Post-intervention agents additionally received the BCRR and, for aggregate/individual profiles, baseline survey responses, reflecting the original study's two-time-point design. Baseline survey responses were taken prior to intervention exposure and hence did not contain post-intervention outcomes. The full list of fields provided to agents is shown in **Multimedia Appendix 1, Tables S1-S4**. Questions were drawn from the Breast Cancer Education Survey and the Risk Report Feedback Survey in the BREATHE study [16]. Four questions were excluded from each survey because they were free text, multiple-response, or used to construct the agents. Response missingness rates were 14.3% for three pre-intervention questions (as limited to participants aged ≥ 45 years) and 8.7% for all questions in the post-intervention survey (due to loss to follow-up).

Unless otherwise stated, n=50 LLM agents were used. Further analysis on larger agent pools showed diminishing returns for this experimental setup.

### 2.2 Simulation of questionnaire responses

Agents were then instructed to complete questionnaires at two timepoints, one before and after the intervention. The pre-intervention questionnaire contained 16 Likert/Categorical questions relating to common breast cancer screening myths, cancer fatalism and screening intentions prior to intervention (**Multimedia Appendix 1, Table S5**). After completing the questionnaire, agents received a BCRR (low, medium or high risk), taken directly from the original study to make the intervention as realistic as possible. For individual-profile agents, BCRR corresponding to their actual risk identified in the BREATHE study were used, whereas age-only, minimal profile and aggregate profile agents were assigned reports according to age-stratified risk distributions. Report information was then added to the agent's memory using the prompt in **Multimedia Appendix 1, Table S6** to mimic how people remember information.

Finally, agents completed the post-intervention survey consisting of 20 questions (all Likert-style; ranging from 5-7 options), which assessed their understanding of BCRR, emotional reactions and intentions to screen after the intervention (**Multimedia Appendix 1, Table S7**). Before answering each question, the memory of BCRR was retrieved and added to the LLM agent's context window.

### 2.3 Prompt template and parameters

We evaluate the use of two different prompt templates (**Multimedia Appendix 1, Table S6**). Each prompt is designed to give the agents questions sequentially. Responses are generated in a stateless manner; i.e. agents answer based on current state without retaining the context of previous questions. This reduces computational load. It also reduces the need to randomize the order of survey questions since the memory effect of prior questions are removed. We avoid presenting multiple questions simultaneously to the LLMs as they may overlook or leave some questions unanswered.

Existing literature [8, 9] has noted a lack of diversity in LLM simulations. To address this, we used soft probability aggregation (**Figure 2**); conceptually similar ideas exist in previous work [17, 18]. Instead of asking the LLM to choose just one answer, we ask it to give a pseudo-probability distribution using numbers, such that the total sum of all options should be 100. The raw scores are then normalized into a probability distribution. If LLMs hallucinate and provide numbers that do not sum to exactly 100, we rescale by the total. We then aggregate all these probability distributions for each question. This approach is referred to as soft probability aggregation (soft voting)**.** By comparison, hard voting assigns each agent to its highest-probability response (argmax) before aggregation. For each question, the LLM returned its reasoning (which may not reflect the model's actual reasoning), probability estimates, and final answer in a structured JSON format. We parsed the raw output directly; constrained decoding was not used to enforce the response format.

For our model hyperparameters, temperature settings were kept to 0.6 unless otherwise mentioned to reflect a balance between response creativity and consistency; although it is worth noting that specific effects can vary on a per-model basis. Top k, top p and min p were kept at model defaults, while context size was kept at 8192 tokens. If an LLM API call failed, the request was automatically retried up to four times using the Python library Tenacity.

"Instruct" models were used; "thinking" modes were disabled. Instead of a physical BCRR, the agents received virtual files generated by converting the PDF files to images. The same LLM model under each evaluation was used to process images and generate text.

**2.4 Simulation assumptions**

As mentioned, demographic variables were sampled independently when generating agents from aggregate data; thus, correlations between demographic characteristics were ignored. Second, agents' responses were collected individually, and agents did not interact with each other throughout the simulation. This is to mimic the original study, which did not require interaction between participants. Third, LLM agents "built from aggregate data" were assigned BCRR according to only age-stratified distributions, while LLM agents built from individual real human profiles were given the reports corresponding to their age and actual assigned risk level. This simplification was made to reduce computational complexity and data requirements. Finally, survey responses are assumed to be valid proxies for underlying screening attitudes.

**2.5 Statistical analysis**

Simulation disagreement can arise from human-human, LLM-LLM and human-LLM discrepancies*.* Because human responses are variable, repeated surveys by the same individual would also provide different responses. Since this was unavailable in our original dataset, we estimated the variation expected from random sampling using 200 bootstrap resamples with replacement from the original dataset for a given agent population (n=50). For each sample, we calculated the difference between the sample and the population ground-truth distribution. These repeated samples therefore provided a reference for the variation expected when estimating a population distribution from a sample of 50 individuals, and we call it "variation from sampling n=50".

We measure disagreement using three metrics for discrete probability distributions: Total Variation Distance (TVD), Jensen-Shannon Distance (JSD) and normalized Wasserstein Distance (NWD). TVD and JSD measure distributional discrepancy, while NWD accounts for the ordinal nature of questions by penalizing larger deviations more heavily than near misses and is thus the preferred choice of metric if available. Thus, NWD is computed only for ordinal questions and not nominal questions. All three metrics are normalized to facilitate comparison across questions, and their ranges are between 0 and 1, with lower values indicating lower distributional errors (higher accuracy). The full formulation is available in **Multimedia Appendix 3.**

**Sensitivity and subgroup analyses**

For robustness, we evaluate the simulation across two LLMs, and report results as mean ± SD over 5 independent runs for each LLM. For our sensitivity analysis, we use the same simulated population across matched runs, changing only the parameter of interest (i.e., temperature, prompt template, etc.). Due to high computational requirements and thus limited runs, sensitivity analyses were evaluated descriptively using the mean Δ change in evaluation metric, standard deviation, and relative percentage change. For subgroup analysis, we used a Linear Mixed Model to compare NWD across demographic subgroups. Because these analyses involved multiple testing across five demographic categories, p-values were corrected using the Holm-Bonferroni method [19]. Topic Modelling and Cosine Similarity were also used to analyze its survey reasoning. Finally, to estimate how survey accuracy changes with LLM agent population size, we pooled agent responses across 5 runs from the post-intervention survey. For each candidate population size *n*, we drew 200 separate random subsamples of *n* agents with replacement from the pooled agent base and calculated the TVD vis-à-vis the ground-truth distribution.

## 2.6 Ethical considerations

This study involved human participants and complied with all relevant institutional and national research ethics guidelines. Ethics approval was obtained from the National Healthcare Group Domain-Specific Review Board, Singapore (reference no 2020/01327; approval date: June 7, 2021). Written informed consent was obtained from participants by trained study coordinators in the participant's preferred language (English, Chinese, or Malay). The informed consent process included permission to use the study data for secondary analyses relevant to BC screening research; therefore, no additional consent was required for the current analysis. Participant privacy and confidentiality were safeguarded throughout the study. All research data were de-identified before analysis, stored on secure servers with restricted access, and handled in accordance with institutional data protection policies to ensure anonymity. No individual participants can be identified in any image in the paper or multimedia appendices. All data and LLM inferences were done on local machines; no information was sent to cloud servers.

## 2.7 Tool Use

We used open weight LLMs from Alibaba's Qwen family (Qwen3.5 9B MLX 4-bit quantized; 23 Mar 2026: https://huggingface.co/mlx-community/Qwen3.5-9B-MLX-4bit), as well as Google's Gemma family (Gemma 4 E4B-IT MLX 4-bit quantized; 10 Apr 2026: https://huggingface.co/lmstudio-community/gemma-4-E4B-it-MLX-4bit). LLM System Prompt Jinja Templates were not modified. LLMs were accessed by calling the API locally from LM Studio. We used an Apple M1 MacBook Pro (2021 version, Apple M1 Max Chip, 64GB RAM, macOS Tahoe v26.5.2). All analyses were performed in Python 3.13. JSD and NWD were computed using SciPy (v1.17.1). Linear Mixed Models and Holm-Bonferroni Corrections were implemented with the python package statsmodels (v0.14.6).

# Results

**Figure 3A** shows the pre-intervention distributional accuracy of the evaluated agents using TVD (we omit NWD since only 3 out of 16 questions were ordinal-type; the rest were nominal-type). For Qwen3.5 9B, Age-Profile had the lowest TVD (0.122 ± 0.0017) followed by Minimal Profile agents (0.123 ± 0.0057), corresponding to reductions of 66.0% and 65.6% relative to the random baseline (0.359 ± 0.016). For Gemma 4 E4B, the Individual Profile agent achieved the lowest TVD (0.157 ± 0.0035), followed by the Aggregate Profile agent (0.161 ± 0.0024). For comparison, repeatedly sampling n=50 responses from observed participant data gave a TVD of 0.052 ± 0.012, showing the distributional variation expected from sampling alone. Meanwhile, a 50/50 split-sample empirical marginal yields a TVD of 0.010 ± 0.01. In contrast to the post-intervention results, agents with more detailed profiles were not consistently the best-performing approach across both LLMs. Similar trends were observed for JSD. Full results are shown in **Multimedia Appendix 2, Tables S1 - S2**.

Additionally, to examine the effect of profile grounding on agent reasoning, we computed pairwise cosine similarity between agent reasonings for each question. All profile grounding methods were associated with higher cosine similarity compared to zero-shot. For both models, aggregate-profile grounding gave the largest increase in similarity (Gemma 4 E4B: $\beta$=0.077, 95% CI 0.069-0.085, $p<0.001$); (Qwen3.5 9B: $\beta$=0.040, 95% CI 0.033-0.047, $p<0.001$) (**Multimedia Appendix 2, Tables S3 - S4**).

**Figure 3B** shows the post-intervention distributional accuracy of the evaluated agents using two different LLMs, measured using NWD, where lower values indicate better agreement with the observed participant response distributions. Across both LLMs, profile-based agents achieved lower NWD (i.e. lower error/ better performance) compared to the random baseline (0.230 ± 0.012) and zero-shot prompting (Gemma 4 E4B - 0.222±0.005; Qwen3.5 9B - 0.181±0.003).

For both LLMs, Individual Profile agents achieved the lowest NWD, with scores of 0.092 ± 0.005 for Gemma 4 E4B and 0.110 ± 0.006 for Qwen3.5 9B. Relative to the random baseline, these results represent reductions of 60.0% and 52.2%, while reductions relative to zero-shot prompting were 58.5% and 39.2%. However, performance was comparable to that of Aggregate Profile agents, which achieved NWD scores of 0.093 ± 0.006 for Gemma 4 E4B and 0.118 ± 0.003 for Qwen3.5 9B. Compared with the random baseline, Aggregate Profile agents reduced NWD by 59.6% and 48.7%, respectively, and by 58.2% and 34.8% relative to zero-shot prompting. For comparison, repeatedly sampling n=50 responses from the observed participant data gave an NWD of 0.037 ± 0.004. Meanwhile, a 50/50 split-sample empirical marginal yields a NWD of 0.008 ± 0.004. Minimal Profile agents (age, ethnicity, and housing type) also improved over the random baseline and zero-shot but performed similarly to age-only agents. For Gemma 4 E4B, the NWD was 0.124 ± 0.010 for Minimal Profile agents and 0.122 ± 0.005 for age-only agents. For Qwen3.5 9B, the NWD was 0.121 ± 0.005 for Minimal Profile Agents and 0.127 ± 0.010 for age-only agents. Similar trends were observed for TVD and JSD (**Multimedia Appendix 2, Tables S5 - S6**). Unlike pre-intervention results where Qwen3.5 9B had the lowest error, Gemma 4 E4B had the lowest post-intervention survey error of the two LLMs.

We also compare the mean (average) Likert responses produced by each grounding method against the human ground truth (**Multimedia Appendix 2, Figure S1 - S2**). Across both LLMs, Individual and Aggregate grounding shifted average responses in the direction of the ground truth compared to zero-shot prompting. Nevertheless, these responses seem to gravitate toward the midpoint of the scale compared to human ground truth even after demographic grounding. For questions where participants agreed (e.g. feeling motivated, happy etc. Range=2.0-2.3 / 5-point scale), agents scored closer to neutral (Range=2.3-2.8). For items where participants strongly disagreed (e.g., feeling anxious, stressed out; Range=3.7-3.9 / 5-point scale), agents also scored closer to the midpoint (Range=3.1-3.5).

In subgroup analysis (**Table 2**), compared with their respective reference groups and after adjustments for multiple comparisons, higher errors were observed for women aged 55-59 years (β=0.02, 95% CI [0.01, 0.03], $p$=0.033), women who never been employed (β=0.06, 95% CI [0.03, 0.08], $p$<0.001), women residing in private property or other non-public housing (β=0.02, 95% CI [0.01, 0.03], $p$<0.001), and widowed women (β=0.06, 95% CI [0.05, 0.08], $p$<0.001). These estimates should be interpreted as exploratory, as estimates for subgroups with small sample sizes should be taken with caution, i.e. "widowed" (n=4), “never employed” (n=4-12). Non-pooled results for each LLM are shown in **Multimedia Appendix 2**, **Tables S7-S8.**

Prediction accuracy varied across survey question themes and LLM models. For the Gemma 4 E4B model (**Figure 4A**), pre-intervention questions related to prevention and screening knowledge (mean TVD = 0.128) and screening barriers (mean TVD = 0.122) had the lowest prediction errors, whereas cancer fatalism questions showed the highest errors (mean TVD = 0.367). Similarly, for the Qwen3.5 9B model (**Multimedia Appendix 2, Figure S3**), cancer fatalism remained the most challenging category (mean TVD = 0.187), while risk perception had the lowest prediction errors (mean TVD = 0.089).

For the cancer fatalism statement, “*there is not much you can do to lower your chances of dying from breast cancer”*, agent reasoning showed strong disagreement relative to humans, often linking choices to the perceived value of early detection (**Multimedia Appendix 4, Table S1**). Yet, the extent of disagreement on the Likert scale varied across models (mean-Likert score: Gemma: 2.0, Qwen: 2.9, Ground-Truth: 3.1). Similarly, 80% of aggregate-profile agents disagreed with the statement *"Breast cancer screening is expensive,"* compared with 59% in the ground truth data (**Multimedia Appendix 2, Figure S4 - S5**). Rationales generated by agents for this question focused on the prior screening history, socioeconomic profile of the individual, and general availability of subsidies (**Multimedia Appendix 4, Table S2**).

Post-intervention prediction errors were slightly different for the two LLMs. For Gemma 4 E4B (**Figure 4B**), the largest errors were observed for positive emotional reactions (mean TVD =

0.232) and attitudes toward genetics (mean TVD = 0.231). In contrast, Qwen3.5 9B exhibited the highest errors for attitudes toward genetics (mean TVD = 0.357) and understanding of the breast cancer report (mean TVD = 0.323) (**Multimedia Appendix 2, Figure S6**).

**Soft probability aggregation versus single choice prompting**

Across all evaluated LLMs and prompting conditions, soft probability aggregation improved agreement with the observed survey distributions compared to sampling a single response. (**Multimedia Appendix 2, Table S9**). The reduction in NWD prediction error for profile agents ranged from 0.006 (5.0%) to 0.017 (15.7%). Similar reductions were observed for JSD and TVD and across different prompt templates.

**Sensitivity to prompt template**

Changing the prompt template from Variant 1 to Variant 2 resulted in a change in NWD (Mean Δ = 0.008 to -0.007; a change of 7.8% to -4.0%) (**Multimedia Appendix 2, Table S10**). Though all profile-grounded agents (mean NWD: 0.1158 - 0.1233) continued to outperform the zero-shot agents (mean NWD = 0.1743), the relative rankings changed (e.g. Individual < Aggregate < Minimal < Age Only < Zero-Shot to Minimal < Individual < Age Only < Aggregate < Zero-Shot) based on NWD. Similar variations were seen for JSD and TVD.

**Sensitivity to temperature hyperparameter**

When the temperature hyperparameter was changed from 0.6 to 0.0 on the Qwen3.5 9B Model, NWD generally increased (Mean Δ NWD ranged from -0.004 to 0.014; a change of -3.3% to 10.5%) (**Multimedia Appendix 2, Table S11**). Nevertheless, all profile-grounded agents (mean NWD: 0.1194 - 0.1359) continued to have better accuracy than zero-shot agents (mean NWD= 0.2215), although the most accurate configuration changed from Minimal Profile Agents to Aggregate Profile Agents.

**LLM differences**

The Gemma 4 E4B had lower NWD than Qwen3.5 9B for both Aggregate and Individual Profile agents, although inference times were longer (~11000s vs ~6000s per run). This is partially attributable to the longer response length for the Gemma model for the same prompt, and their sample outputs are shown in **Multimedia Appendix 4, Table S3.**

**Relationship between errors and number of simulated agents**

Discrepancies between simulated and observed response distributions simulated with soft-probability aggregation decreased rapidly with increasing agent population size, reaching a plateau at approximately 50-100 agents (**Multimedia Appendix 2, Figure S7**).

# Discussion

**4.1 Principal results**

Previous studies have explored using LLMs to simulate engagement with maternal health messaging interventions [20], vaccination decisions [21, 22] and symptom reporting behavior during a pandemic [23]. Our research extends this literature. Using data from a retrospective study, we found that profile grounding improved survey fidelity relative to a zero-shot prompt. Aggregate and Individual Profile agents (compared to age and minimal profiles) achieved the highest post-intervention performance. However, the limited gains from demographic attributes alone suggest that culturally relevant factors may be important for reproducing survey responses. The effect of profile grounding appears to depend on the model, prompt used, grounding method, and model hyperparameters. These findings support the potential of profile-grounded LLMs for *in silico* simulation while highlighting the importance of cultural context and sensitivity to model and prompting choices.

#### 4.1.1 Profile grounding improved survey fidelity relative to zero-shot approach

The benefit of providing contextual knowledge of participants over zero-shot prompting suggests that LLM agents require contextual anchoring to approximate human survey respondents. That aggregate and individual profile agents performed best post-intervention, while simpler profiles (age-only, minimal) sufficed pre-intervention for one model, suggests that baseline health beliefs may be sufficiently captured by broad demographic signals, whereas responses to a personalised health intervention demand richer, individual-level context. This asymmetry implies that the appropriate level of grounding is not a fixed design parameter but should be matched to the complexity of the construct being simulated. The modest difference in performance between aggregate and individual profiles has practical implications given the computational and data-access costs of individual-level profiling at scale.

Notably, repeated sampling of only 50 participants produced lower errors than the LLM-based approaches we tested. This suggests that aggregate response distributions can be approximated well with a small subset of the population. However, LLM simulations may still offer value if collecting even small amounts of real-world data is impractical.

#### 4.1.2 Profile grounding effects raise questions about what LLMs are simulating

The greater cross-model consistency of profile grounding effects post-intervention raises a question about what LLMs are learning to simulate. One interpretation is that personalised health interventions make individuals' responses more closely aligned with demographic information, making the agent's task more manageable regardless of model architecture. Another is that post-intervention states involve stronger social desirability or stereotype-consistent responding, which LLMs replicate without representing genuine individual variation. Understanding which mechanism is operating will matter for validity.

Profile grounding was also associated with increased reasoning similarity and longer response lengths. However, agents continued to favour midpoint Likert responses even after grounding. Increased reasoning similarity may reflect appropriate use of task-relevant demographic priors (i.e. agents anchor on provided context rather than generating generic responses). The same mechanism, however, may suppress within-group diversity. When models rely heavily on demographic characteristics, they may reduce the heterogeneity present within real demographic groups [8, 10]. The resulting distributions may then be too narrow to capture the full range of opinions from a real survey. The persistence of midpoint responding adds to this concern. Whether this reflects an "average persona" effect [24] or a limitation in how LLMs map graded attitudes onto ordinal scales, the consequence is the same. Simulated distributions will underrepresent both strong supporters and strong resisters, biasing downstream analyses that depend on variance. Future work should examine whether richer affective prompting or calibration against empirical distributions can restore the spread that demographic grounding alone does not provide.

#### 4.1.3 LLM simulation accuracy varies across survey constructs and sociodemographic subgroups

Similar to other studies investigating synthetic survey generation, we found that LLMs reproduced some population-level response distributions more effectively than others [10, 21, 25]. Questions relating to screening knowledge and intentions were associated with relatively low error rates, whereas fatalistic beliefs and genetic attitudes showed higher errors across both LLMs, suggesting that simulation accuracy varies with the underlying survey construct. This pattern may reflect the degree to which constructs are grounded in culturally situated beliefs. Fatalism and genetic attitudes are shaped by deeply held values that vary across ethnic and social contexts and are less likely to be well-represented in LLM training material that skew toward Western, English-language sources [10, 26].

Prediction errors were also higher for topics such as perceived affordability of breast cancer screening, as LLMs rated screening as less expensive than human participants. Topic modelling for the question showed that agent reasonings often invoked respondents' prior screening history, socioeconomic profile, and the availability of government subsidies, leading them to overconfidently infer that cost was unlikely to be a major barrier. LLMs may have overgeneralised from aggregate knowledge of Singapore's healthcare system and underweighted the lived financial constraints reported by participants.

At the demographic level, we observed statistically significant increases in prediction errors among women aged 55-59 and those living in private properties. Given the stochasticity of LLM simulations and the limited scope of the questionnaire, the specific subgroups causing this difference should be interpreted with caution. Nevertheless, these findings are consistent with a growing literature reporting that LLM predictions exhibit biases across demographic subgroups [7, 8, 10, 15, 21, 27, 28]. Profile grounding alone may be insufficient to fully capture the variation in subjective attitudes and beliefs that are shaped by lived experience, cultural context, and community norms.

Together, these findings highlight that simulation fidelity is uneven in ways that are not random but influenced by cultural distance and demographic representativeness. This has direct implications for how LLM-based simulations should be validated and deployed in public health research. Future work should explore localisation strategies that include supervised fine-tuning and bias rectification [29-31]. It remains unclear whether their effectiveness extends across diverse populations and beyond constrained tasks such as missing data imputation to more complex use-cases such as population response forecasting. Subgroup- and construct-level evaluation should thus be treated as a minimum standard before LLM simulations are used to inform screening policy or behaviour change interventions.

#### 4.1.4 Variation in simulation outputs across models, prompts, and temperature settings

LLM simulation outputs tend to be stable within a fixed configuration. In our experiments, repeated runs using the same model, prompt template, and temperature generally produced small standard deviations (*low within-configuration variance*). However, changing any of these parameters came with interpretive consequences and affected overall simulation accuracy and occasionally the relative performance of different profile-grounding strategies (*high between-configuration variance*). For example, reducing temperature from 0.6 to 0.0 generally worsens our simulation accuracy. Meanwhile changing prompt wording sometimes improves it. These choices affected not only overall accuracy across all profile agents, but also their relative rankings.

#### 4.1.5 Demographic prompting effectiveness as a function of model and task characteristics

This sensitivity to configuration extends to the design of profile grounding itself. Kamruzzaman et al. [32] found that the effectiveness of demographic prompting depends on model architecture, task characteristics and the quality of the demographic attribute signals. Hence, additional demographic attributes sometimes degrade performance if these signals conflict. This may explain why some studies report successful performance with demographic profiles, while others do not [15]. Additionally, a simulation's result is conditional on both the configuration that the researcher chooses and the treatment being implemented. The implication is that studies examining different parts of this configuration space may reach different conclusions.

#### 4.1.6 Balancing simulation complexity against empirical validation

Our findings also relate to an issue about the "correct" level of complexity in LLM-based simulations [33, 34]. More advanced architectures [35, 36] have incorporated components such as planning, reflection, social interaction, and dynamic internal states. These methods might produce more realistic behaviour; yet they also introduce additional parameters that are harder to validate [33, 37]. While we tried to adopt a simpler design based on demographic information and baseline survey responses, performance remained sensitive to basic implementation choices, suggesting that the desire to increase simulation complexity must be balanced with empirical validation.

### 4.2 Limitations

Our study is subject to several limitations that should be considered when interpreting the findings.

First, simulation performance is contingent on a range of implementation choices, including agent construction, prompt design, and output formatting. Hence, observed performance may reflect not only limitations of the LLMs used but also assumptions embedded in the simulation code. Moreover, small changes in prompt phrasing and output formatting produced notable differences. This variance is unfortunately not captured by the reported standard deviations, which only reflect differences across simulation runs rather than the possibly infinite sample space of prompt formulations.

Second, a related concern involves the statistical properties of LLM-generated response distributions. Prior work has demonstrated that such distributions tend to exhibit lower variance than human response distributions under equivalent conditions [24]. This compression of variance inflates standardised effect sizes and undermines the assumptions of conventional significance tests, causing treatment effects to appear larger than they are in practice [29]. Simulation-based findings should therefore not be interpreted using the same inferential standards applied to human participant data without first validating that simulated and human response variances are comparable [18].

Third, comparing pre- and post-intervention survey accuracy should be done carefully, since the two questionnaires were not equivalent in scope, question constructs, or response formats. Observed differences in simulation accuracy across time points may therefore partly reflect measurement artefacts attributable to questionnaire design rather than genuine changes in agent predictive validity.

Fourth, profile agents were constructed exclusively from demographic variables and baseline survey responses. Many other real-world factors such as health literacy, human personality and cultural context were not available to the agents; and possibly difficult to implement without introducing researcher bias. Moreover, simulations were conducted in English. LLM Prompt language effects on our simulation remain unknown, although evidence that it helps improve cultural alignment remains mixed [11, 38]. The omission of these factors likely placed an upper bound on achievable simulation accuracy and may have introduced systematic error if the omitted factors are differentially distributed across the demographic groups represented in the cohort.

Future simulation work requires even greater consideration not only for the ways that agents are constructed but also how they are evaluated. Since surveys were used as validation and benchmarking tools, a survey with limited scope negatively impacts the quality of inferred findings [39]. A possible extension would be to integrate more quantitative and qualitative factors of human behavior reproducibility [37].

### 4.3 Conclusions

Our simulation explores individual and aggregated data profile grounding to predict population-level responses post-intervention. Although LLM simulations are not a substitute for human participants, they might still be used to explore potential responses and insights following public health interventions. To achieve this, future work can look towards quantifying simulation uncertainty and integrating local / context-specific knowledge to address uneven performance across sociodemographic groups and health attitudes / beliefs.

## ACKNOWLEDGEMENTS

We sincerely acknowledge the work of Park et al. [6], whose codebase formed the initial development of this research. It was then modified to fit the requirements of this study. Any issues remain solely the responsibility of the authors of this paper.

**FUNDING STATEMENT**
This study was funded by the JurongHealth Fund (reference no JHF-20-RE-003), Agency for Science, Technology and Research (A*STAR), and the Precision Health Research Singapore Clinical Implementation Pilot (PRECISE CIP) Fund. MH is supported by the JurongHealth Fund, PRECISE CIP Fund, the Breast Cancer Prevention Programme under the Saw Swee Hock School of Public Health Programme of Research Seed Funding (SSHSPH-Res-Prog-BCPP), the Breast Cancer Screening Prevention Programme under the Yong Loo Lin School of Medicine (NUHSRO/2020/121/BCSPP/LOA), the National University Cancer Institute Singapore (NCIS) Centre Grant Programme (CGAug16M005), and the Asian Breast Cancer Research Fund. The funders were not involved in the study design, data collection, analysis, interpretation, or the writing of the manuscript. KK is a recipient of the A*STAR Research Internship Award (ARIA).

**CONFLICTS OF INTEREST**
None declared.

**DATA AVAILABILITY**
BREATHE individual-level data are owned by the providing institutions (Ng Teng Fong General Hospital [NTFGH], National University Hospital [NUH], Alexandra Hospital [AH], National University Polyclinics [NUP], and Jurong Medical Centre [JMC]). Data may be obtained upon a reasonable request to the principal investigator, MH (ephbamh@nus.edu.sg). The data are not publicly available due to privacy and/or ethical restrictions. Legal agreements will need to be drawn up between data requesters and providers for access to the de-identified data. The proposed studies need to comply with Singapore's laws and regulations regarding human biomedical research and clinical investigation, including the Declaration of Helsinki, International Good Clinical Practice Guidelines, and Good Clinical Practice guidelines by Singapore's Health Sciences Authority and the Ministry of Health.

**CODE AVAILABILITY**
The code used to run the simulation and perform all analyses are available at:
https://github.com/andantemoss/LLM_BC_Simulation_Intervention

This repository includes the simulation code, analysis scripts, prompt templates and experiment configuration files. It also specifies the random seeds, software dependencies and procedures to create the experiments. Sample simulated data inputs have been provided in place of actual data used for this research due to restrictions on data sharing.

**AUTHOR CONTRIBUTIONS**
Conceptualisation: JL
Methodology: KK, RJYL, AS, BLD, ML
Data curation: MH, PJH
Formal analysis: KK
Writing - original draft: KK

Writing - review and editing: All
Visualisation: KK
Supervision: JL

**ABBREVIATIONS**

| | |
|---|---|
| API | Application Programming Interface |
| BC | Breast Cancer |
| BCRR | Breast Cancer Risk Report |
| LLM | Large Language Models |
| JSD | Jensen-Shannon Distance |
| TVD | Total Variation Distance |
| NWD | Normalized Wasserstein Distance (also known as Earth Mover Distance) |
| IPF | Iterative Proportional Fitting |
| SD | Standard Deviation |
| SE | Standard Error |
| PDF | Portable Document Format by Adobe |

**DECLARATION OF AI USE**

With the assistance of Microsoft 365 Copilot, portions of the text in this manuscript were edited for brevity, as well as to improve the matplotlib (Python Library) code used to generate some figures. All LLM outputs were reviewed and edited, and we thus take full responsibility for all final content, analysis, tables, and figures shown in this manuscript.

## Table 1. Methods compared

| Type | Method | Demographic Variables | Intervention given [a] | Baseline Attitudes | Data Requirement(s) |
|---|---|---|---|---|---|
| Statistical | Uniform Random Baseline | ✗ | ✗ | ✗ | None |
| | Majority Response Baseline | ✗ | ✗ | ✗ | Per Question Response Frequency Distributions |
| | Variation from sampling n=50 | ✗ | ✗ | ✗ | Random samples of 50 responses from the full survey dataset |
| | 50/50 Split Empirical Marginal | ✗ | ✗ | ✗ | Full Survey Responses (split 50/50) |
| LLM-Based | Zero-Shot | ✗ | ✗ | ✗ | None |
| | Age only Agents | Age | ✓ | ✗ | Age Frequency Distribution |
| | Minimal Profile Agents | Age, Ethnicity, Housing Type | ✓ | ✗ | Age, Ethnicity, Housing Type Frequency Distribution |
| | Aggregate Profile Agents | ✓ | ✓ | ✓ | Aggregated Survey Data |
| | Individual Profile Agents | ✓ | ✓ | ✓ | Individual-Level Survey Data |

[a] Intervention report requires age to be provided.

## Table 2. Prediction errors across demographic subgroups

| Category | Subgroup | Pooled Agents per run, n [a] | Mean NWD (SD) | β Coefficient[b] | Estimate [b] (95% CI) | Uncorrected P value [c] | Corrected P value [d] |
|---|---|---|---|---|---|---|---|
| **Age Group** | 35-39 | 44-52 | 0.0746 (0.035) | Reference | Reference | - | - |
| | 40-44 | 88-94 | 0.0750 (0.041) | 0.000405 | 0.00 (-0.01 to 0.01) | 0.946 | 1.00 |
| | 45-49 | 84-94 | 0.0761 (0.037) | 0.001512 | 0.00 (-0.01 to 0.01) | 0.799 | 1.00 |
| | 50-54 | 80-104 | 0.0796 (0.047) | 0.005022 | 0.01 (-0.01 to 0.02) | 0.398 | 1.00 |
| | 55-59 | 74-86 | 0.0922 (0.048) | 0.017661 | 0.02 (0.01 to 0.03) | 0.003 | 0.033 |
| **Employed** | Yes | 326-354 | 0.0685 (0.039) | Reference | Reference | - | - |
| | Never employed | 4-12 | 0.132 (0.097) | 0.056479 | 0.06 (0.03 to 0.08) | <0.001 | <0.001 |
| | Previously employed/retired | 42-62 | 0.0665 (0.034) | -0.00856 | -0.01 (-0.03 to 0.02) | 0.508 | 1.00 |
| **Ethnicity** | Chinese | 306-308 | 0.0768 (0.040) | Reference | Reference | - | - |
| | Indian | 22-40 | 0.0814 (0.043) | 0.004662 | 0.00 (-0.01 to 0.02) | 0.427 | 1.00 |
| | Malay | 42-58 | 0.0719 (0.035) | -0.004914 | -0.00 (-0.02 to 0.01) | 0.402 | 1.00 |
| | Others | 12 | 0.0890 (0.043) | 0.012265 | 0.01 (0.00 to 0.02) | 0.037 | 0.366 |
| **Housing Type** | HDB 1-4 room | 162-176 | 0.0713 (0.039) | Reference | Reference | - | - |
| | HDB Executive or 5-room flat | 134-146 | 0.0729 (0.040) | 0.001621 | 0.00 (-0.01 to 0.01) | 0.703 | 1.00 |
| | Private Property and Others | 90-92 | 0.0891 (0.043) | 0.017795 | 0.02 (0.01 to 0.03) | <0.001 | <0.001 |
| **Marital Status** | Currently Married | 306-324 | 0.0726 (0.039) | Reference | Reference | - | - |
| | Never Married | 42-58 | 0.0802 (0.039) | 0.007489 | 0.01 (-0.01 to 0.02) | 0.254 | 1.00 |
| | Divorced | 32-34 | 0.0812 (0.046) | 0.008553 | 0.01 (-0.00 to 0.02) | 0.193 | 1.00 |
| | Widowed | 4 | 0.133 (0.082) | 0.062547 | 0.06 (0.05 to 0.08) | <0.001 | <0.001 |

*[a] Pooled agents per run reflect the sample pooled across both LLMs within each run (200 agents per LLM; 400 agents total per run). Two runs were conducted, with ranges reflecting variation in subgroup counts across runs. More agents were used for this analysis to ensure adequate representation of smaller demographic subgroups. Refer to Multimedia Appendix 2, Table S7-S8 for the non-pooled results, which report results separately for each LLM model.*

*[b] Estimated change in NWD relative to the reference subgroup.*

*[c] Linear Mixed Effects Model (**Multimedia Appendix 3**) included demographic category as a fixed effect; 'question' (question number) and 'run' (run 1, run 2) were included as random effects to account for repeated questions and run to run variability. Individual profile agents were used as it allows comparison of subgroup-specific NWDs with the corresponding subgroup ground-truth distributions.*

*[d] Corrected via Holm-Bonferroni Correction. To control family-wise error rates across multiple demographic categories representing multiple hypothesis, raw p-values from reference group comparisons across all demographic variables were pooled into a single vector and adjusted. The correction was done using statsmodels library (v0.14.6) using the multipletests function with the argument method='holm'*

**Figure 1. Simulation design**

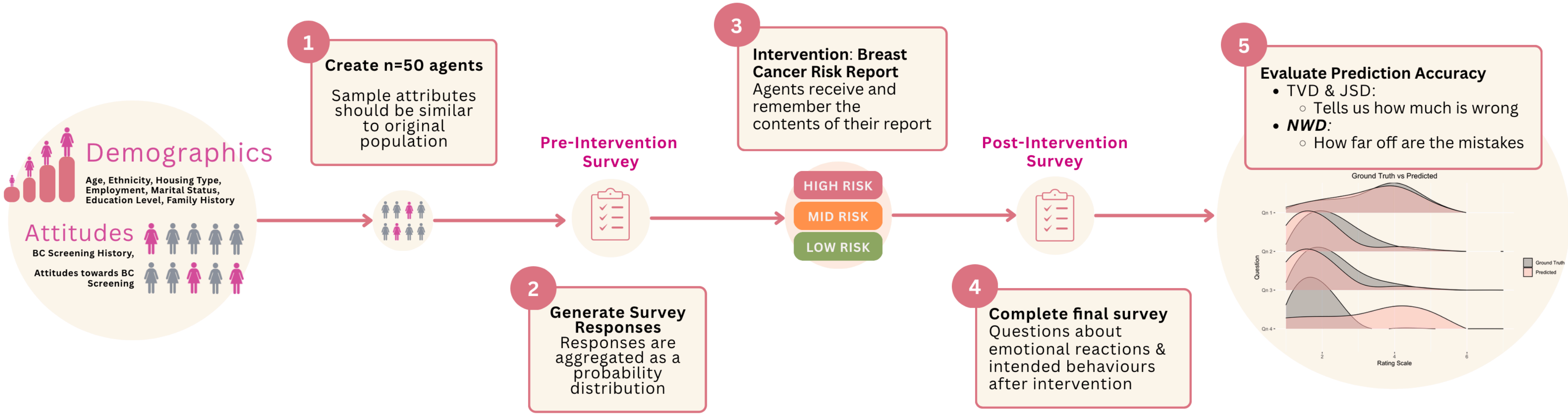

**Figure 2:** Illustration of soft probability aggregation vs hard voting

**Question:** Breast Cancer Screening is expensive

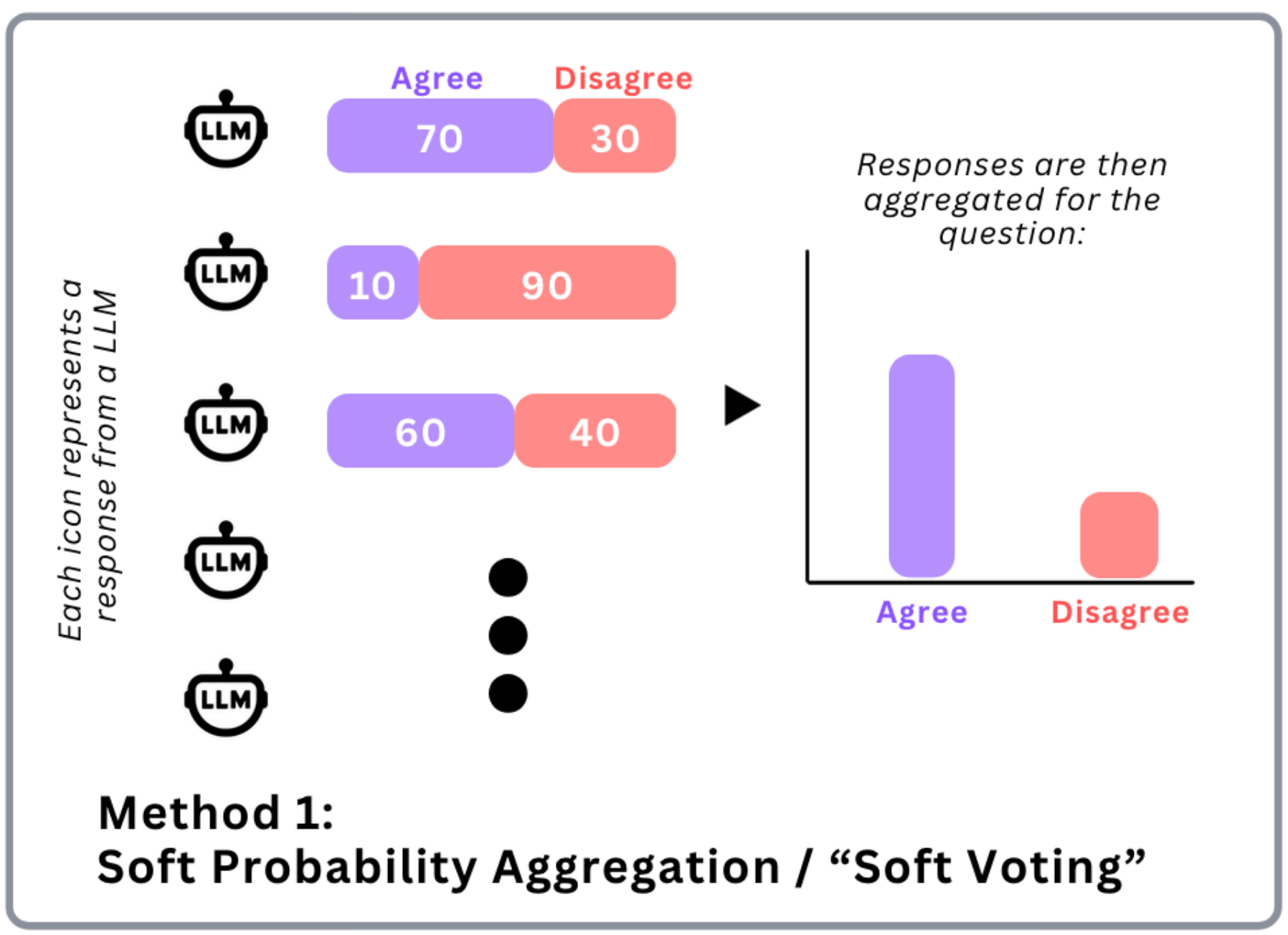

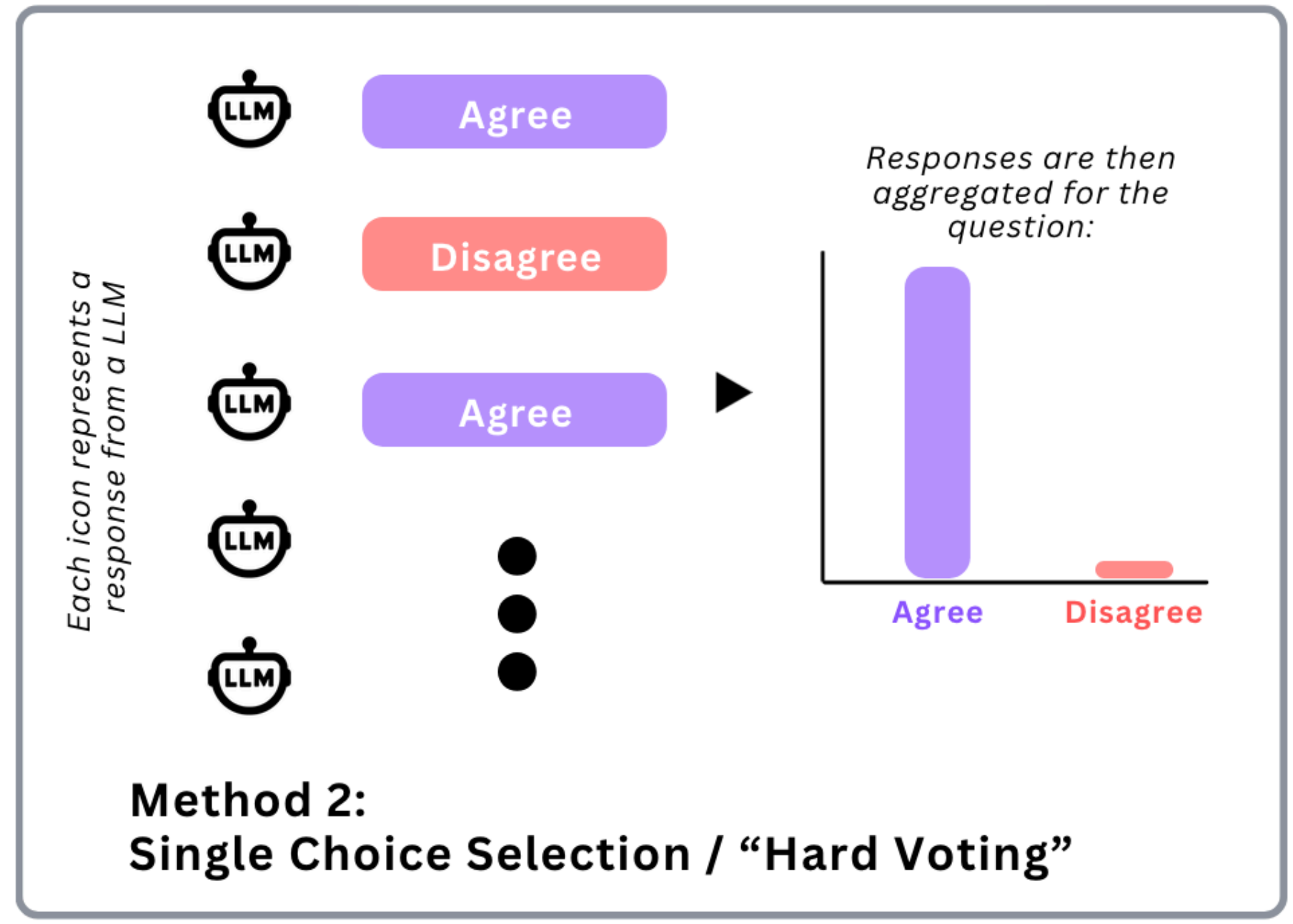

**Figure 3.** Survey distributional accuracy pre- (A) and post-intervention (B).

TVD was the chosen metric at the pre-intervention time point, as some questions were nominal-type. NWD was the chosen metric for the post-intervention time point, as all questions were ordinal type (Likert-scale). *Lower values are better (i.e. lower error) for both metrics.*

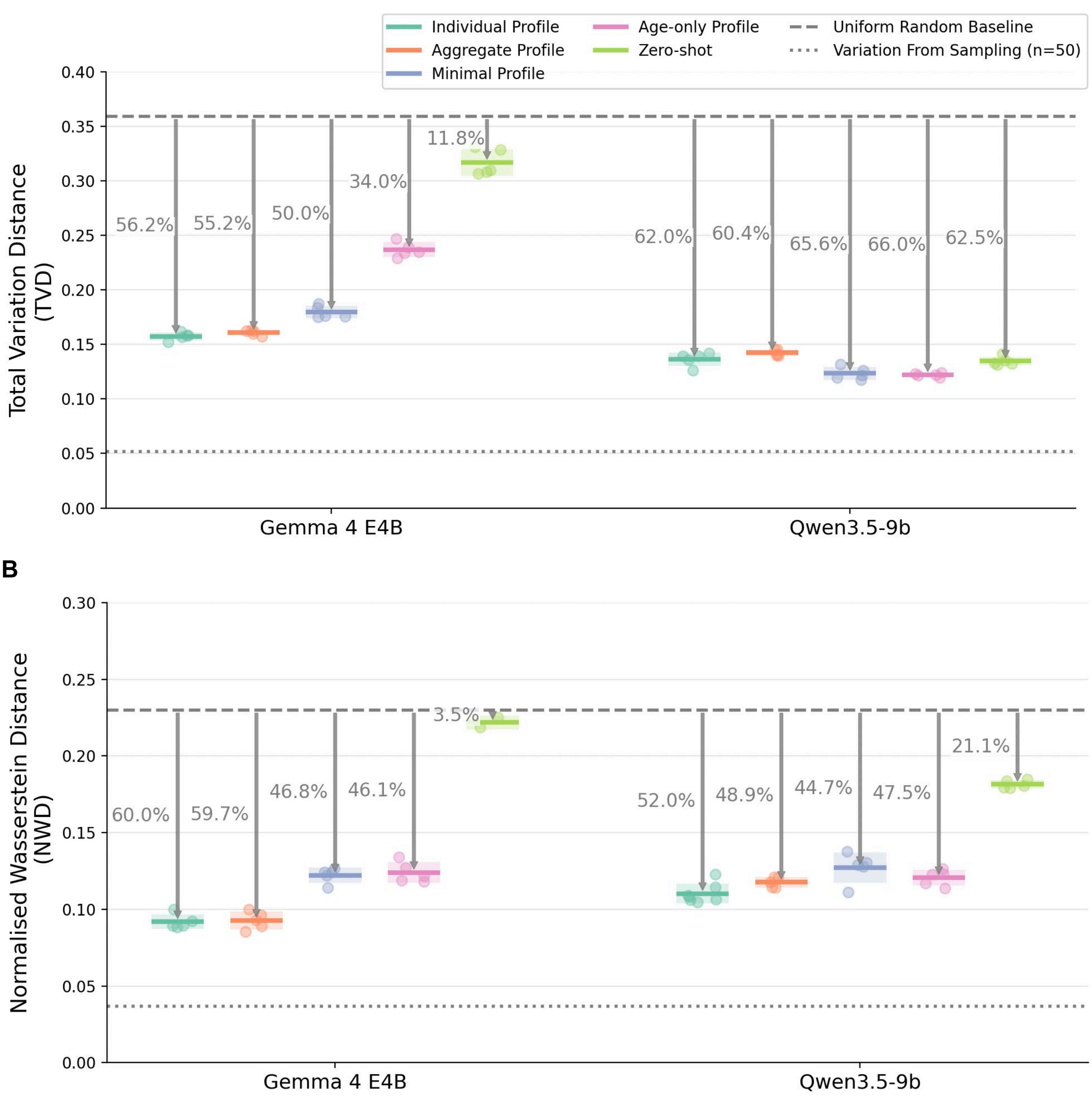

**Figure 4. Prediction errors by question theme (A) pre- and B) post-intervention.** Computed using Gemma 4 E4B (Aggregate Profile Agent): Heatmap cells display the means for TVD, JSD and NWD with standard deviations across 5 simulation runs. Color represents the relative standing for a given metric; green becomes red as error increases.

***A:*** *Pre-Intervention*

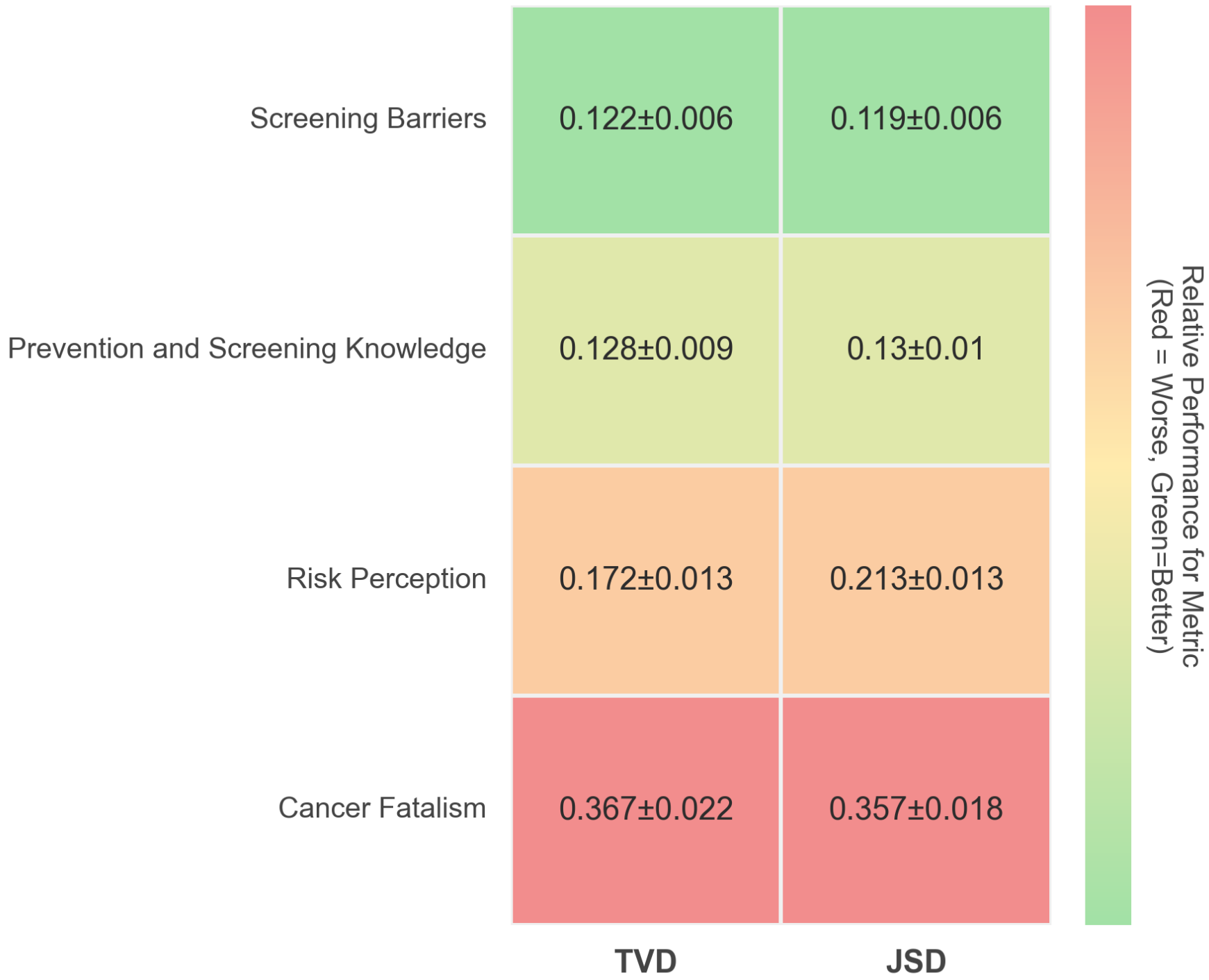

***B****: Post-Intervention*

| | TVD | JSD | NWD |
|---|---|---|---|
| Behavioral Intentions after Intervention | 0.093±0.023 | 0.105±0.021 | 0.035±0.007 |
| Understanding of Breast Cancer Report | 0.17±0.013 | 0.227±0.013 | 0.077±0.009 |
| General Well Being | 0.208±0.019 | 0.229±0.018 | 0.078±0.01 |
| Negative Emotional Reactions | 0.217±0.024 | 0.225±0.02 | 0.12±0.011 |
| Perceived Breast Cancer Risk after Intervention | 0.221±0.017 | 0.236±0.024 | 0.072±0.009 |
| Attitudes towards genetics | 0.231±0.013 | 0.252±0.01 | 0.103±0.008 |
| Positive Emotional Reactions | 0.232±0.037 | 0.261±0.029 | 0.128±0.019 |

Relative Performance for Metric
(Red = Worse, Green = Better)

## MULTIMEDIA APPENDIX

**Multimedia Appendix 1.** Simulation inputs (Agent profile fields, questionnaires and prompt templates)
**Multimedia Appendix 2.** Detailed tables and figures for experimental results
**Multimedia Appendix 3.** Distributional similarity metrics used for evaluating survey accuracy against human response distributions
**Multimedia Appendix 4.** Simulation outputs (sample outputs for each LLM Model, as well as topic modelling results on LLM Agent reasonings for chosen questions)

**Multimedia Appendix 1 – Simulation Inputs**

**Table S1.** Pre- and Post-Intervention Fields for Age-only Profile Agents

| Variable | Type | Levels used in analysis | Category |
|---|---|---|---|
| Age | Continuous | Years | Demographics |

**Table S2.** Pre- and Post-Intervention Fields for Minimal Profile Agents

| Variable | Type | Levels used in analysis | Category |
|---|---|---|---|
| Age | Continuous | Years | Demographics |
| Ethnicity | Categorical | Chinese, Malay, Indian, Other | |
| Housing type | Categorical | HDB 1–3 room, HDB 4-room, HDB 5-room, HDB Executive, Private condominium, Landed property, Other | |

**Table S3.** Pre-Intervention Fields for Aggregate & Individual Profile Agents

| Variable | Type | Levels used in analysis | Category |
|---|---|---|---|
| Age | Continuous | Years | Demographics |
| Ethnicity | Categorical | Chinese, Malay, Indian, Other | |
| Housing type | Categorical | HDB 1–3 room, HDB 4-room, HDB 5-room, HDB Executive, Private condominium, Landed property, Other | |
| Marital status | Categorical | Never married, Married, Widowed, Separated/Divorced | |
| Employment status | Categorical | Currently/previously employed or retired, Never employed | |
| Smoking status | Categorical | Ever smoker, Never smoker | |
| Alcohol consumption | Categorical | Drinking frequency (times/year) | |
| Health for the past week | Categorical | Very poor, Poor, Fairly poor, Average, Fairly good, Good, Excellent | |
| Quality of life for the past week | Categorical | Very poor, Poor, Fairly poor, Average, Fairly good, Good, Excellent | |
| Family history of breast cancer | Categorical | Yes, No, Do not know | |
| Occupation | Categorical | Business & Management, Sales & Service, Healthcare, Education, Finance, Manufacturing & Logistics, Government & Public Service, Technology, Other, Administration, Skilled Trades, Engineering & Technical, Research & Development, Operations & Management, Retail & Sales, Clerical & Administrative, Cleaning & Maintenance, Healthcare & Research, Construction & Engineering | |
| Highest educational qualification | Categorical | No formal education; Primary; Secondary; Pre-university; (ITE/NITEC/NTC); Diploma; Bachelor's degree; Postgraduate | |
| Breast screening attendance | Categorical | No, Yes (Once a year, Once every 2 years, Do not intend to go anymore, Others) | Baseline Attitudes |
| Importance of breast cancer screening | Categorical | Strongly Disagree, Disagree, Neither Agree nor Disagree, Agree, Strongly Agree | |

**Table S4.** Post-Intervention Fields for Aggregate & Individual Profile Agents

| Variable | Type | Levels used in analysis | Category |
|---|---|---|---|
| Age | Continuous | Age in years | Demographics |
| Ethnicity | Nominal | Chinese, Malay, Indian, Other | |
| Housing type | Ordinal | HDB 1–3 room, HDB 4-room, HDB 5-room, HDB Executive, Private condominium, Landed property, Other | |
| Marital status | Nominal | Never married, Married, Widowed, Separated/Divorced | |
| Employment status | Nominal | Employed; Previously employed/retired; Never employed | |
| Smoking status | Binary | Ever smoker; Never smoker | |
| Alcohol consumption | Ordinal | Drinking frequency (times/year) | |
| Family history of breast cancer | Nominal | Yes; No; Unknown | |
| Occupation | Nominal | Business & Management, Sales & Service, Healthcare, Education, Finance, Manufacturing & Logistics, Government & Public Service, Technology, Other, Administration, Skilled Trades, Engineering & Technical, Research & Development, Operations & Management, Retail & Sales, Clerical & Administrative, Cleaning & Maintenance, Healthcare & Research, Construction & Engineering | |
| Highest educational qualification | Ordinal | No formal education; Primary; Secondary; Pre-university; (ITE/NITEC/NTC); Diploma; Bachelor's degree; Postgraduate | |
| Breast screening attendance | Nominal | No, Yes (Once a year, Once every 2 years, Do not intend to go anymore, Others) | Baseline Attitudes |
| Importance of breast cancer screening | Ordinal (5-point Likert) | Strongly Disagree, Disagree, Neither Agree nor Disagree, Agree, Strongly Agree | |
| Perceived chance of getting breast cancer | Ordinal (7-point Likert) | 1 (Lowest) -- 7 (Highest) | |
| Fatalistic belief: not much you can do to lower your chances of getting breast cancer | Ordinal (7-point Likert) | 1 (Strongly disagree) -- 7 (Strongly agree) | |
| Fatalistic belief: not much you can do to lower your chances of dying from breast cancer | Ordinal (7-point Likert) | 1 (Strongly disagree) -- 7 (Strongly agree) | |
| Risk perception: I do not have any family history of breast cancer therefore I will not get breast cancer | Binary | Agree; Disagree | |
| Risk perception: I am still young therefore I | Binary | Agree; Disagree | |

| | | |
|---|---|---|
| will not get breast cancer | | |
| Risk perception: I am still young therefore I do not need to screen for breast cancer | Binary | Agree; Disagree |
| Risk perception: I have given birth therefore I will not get breast cancer | Binary | Agree; Disagree |
| Risk perception: I have breastfed therefore I will not get breast cancer | Binary | Agree; Disagree |
| Prevention knowledge: I can prevent myself from getting breast cancer by eating healthily exercising not drinking alcohol and not smoking | Binary | Agree; Disagree |
| Screening knowledge: If breast cancer is detected early chances of surviving is high | Binary | Agree; Disagree |
| Screening barrier: Breast cancer screening is embarrassing | Binary | Agree; Disagree |
| Screening barrier: Breast cancer screening is expensive | Binary | Agree; Disagree |
| Screening knowledge: Mammogram can trigger cancer cells to grow | Binary | Agree; Disagree |

**Table S5.** Pre-Intervention Questionnaire

| Question | Missingness Count | Question Theme | Question Type |
|---|---|---|---|
| What do you think is your chance of getting breast cancer (`1` = lowest chance; `7` = highest chance) | 0 (0%) | Risk Perception | Likert-Scale (Ordinal) Question |
| Would you say you agree that there is not much you can do to lower your chances of getting breast cancer (`1` = strongly disagree; `7` = strongly agree) | 0 (0%) | Cancer Fatalism | |
| Would you say you agree that there is not much you can do to lower your chances of dying from breast cancer (`1` = strongly disagree; `7` = strongly agree) | 0 (0%) | Cancer Fatalism | |
| I do not have any family history of breast cancer therefore I will not get breast cancer (Agree / Disagree) | 0 (0%) | Risk Perception | Nominal (Categorical) Question |
| I am still young therefore I will not get breast cancer (Agree / Disagree) | 0 (0%) | Risk Perception | |
| I am still young therefore I do not need to screen for breast cancer (Agree / Disagree) | 0 (0%) | Risk Perception | |
| I have given birth therefore I will not get breast cancer (Agree / Disagree) | 0 (0%) | Risk Perception | |
| I have breastfed therefore I will not get breast cancer (Agree / Disagree) | 0 (0%) | Risk Perception | |
| I can prevent myself from getting breast cancer by eating healthily, exercising, not drinking alcohol and not smoking (Agree / Disagree) | 0 (0%) | Prevention and Screening Knowledge | |
| If breast cancer is detected early chances of surviving is high (Agree / Disagree) | 0 (0%) | Prevention and Screening Knowledge | |
| Breast cancer screening is embarrassing (Agree / Disagree) | 0 (0%) | Screening Barriers | |
| Breast cancer screening is expensive (Agree / Disagree) | 0 (0%) | Screening Barriers | |
| Mammogram can trigger cancer cells to grow (Agree / Disagree) | 0 (0%) | Prevention and Screening Knowledge | |
| It is inconvenient for me to go for a mammogram. (Agree / Disagree) | 645 (14.3%) | Screening Barriers | |
| I must have symptoms first before I decide to go for mammogram. (Agree / Disagree) | 645 (14.3%) | Prevention and Screening Knowledge | |
| I rather go for breast ultrasound because mammogram is painful. (Agree / Disagree) | 645 (14.3%) | Screening Barriers | |

**Table S6. Prompt Templates**

**Prompt Template – Variant 1**

Variables
!<INPUT 0>!

=====

Assume you are the participant doing the survey in Singapore.
Answer from your own perspective, using only the beliefs, experiences, attitudes, and circumstances described above. Focus on the factors most relevant to the question.

Provide:

A brief interpretation of the response scale.
A short explanation (2–5 sentences) for your predicted answer.
A single predicted response.

Requirements:
- ResponseDistribution must contain every response option provided in the question.
- Weights must be out of 100 and sum to 100.
- Response should be the option with the highest weight.

Question:

!<INPUT 1>!

---

Output ONLY valid JSON.

{
"Q": "<repeat the question you are answering>",
"Reasoning": "<reasoning on which option makes the most sense for you>"
"ResponseDistribution": {
<for EACH option in the question, include an entry "option_value": weight. Sum of all weights should be 100>
},
"Response": <a single !<INPUT 2>! value that best represents your prediction on how the participant's answer>
}

**Prompt Template – Variant 2**

Variables:

<commentblockmarker>###</commentblockmarker>
!<INPUT 0>!

=====
Task:

Assume you are the woman doing the survey, using only the beliefs, experiences, attitudes, and circumstances described above, based in Singapore.
Use fast, intuitive reasoning. Do not overthink.

Requirements:
- "ResponseDistribution" must contain every response option provided in the question.
- Weights must be out of 100 and sum to 100.
- "Response" should be the option with the highest weight.

Question:

!<INPUT 1>!

---

Output ONLY valid JSON.

{
 "Q": "<repeat the question you are answering>",
 "ResponseDistribution": {
 "<option>": <integer weight>,
 ... add as necessary according to options
 },
 "Response": <a single !<INPUT 2>! value that best represents your prediction>,
 "Reasoning": "<reasoning on which option makes the most sense for you>"
}

**Memory Prompt**

Thank you for taking part in BREATHE (Breast Screening Tailored for Her), a study designed to help women understand their personal risk of developing breast cancer. Breast Cancer Risk Report is attached.

---
Above is something the interviewee ("Item 1") experienced.

Task:
1. Rate the usefulness of this knowledge on a scale from 0 to 100, where 0 represents 'not important' and 100 represents 'very important'.
2. Write how a person would recall it, not copying the exact words. Keep it as short as possible, more words if important and less words if not important.

Hint: If the interviewee's answer is correct, the info is of low importance since it does not change the person's understanding.

Output format: JSON dictionary with the following structure:
{
 "Item 1": {
 "importance": <int importance score (0-100)>,
 "summary": "<detailed memory trace>"
 }
}

**Table S7**. Post-Intervention Questionnaire

| • Question | Missingness Count | Question Theme | Question Type |
|---|---|---|---|
| How would you rate your overall health during the past week? | 394 (8.7%) | General Well-Being | Likert-Scale (7 options)<br><br>(`1`=Very poor; `2`=Poor, `3`=Fairly Poor, `4`=Average, `5`=Fairly Good, `6`=Good, `7`=Excellent) |
| How would you rate your overall quality of life during the past week? | 394 (8.7%) | General Well-Being | Likert-Scale (7 options) |
| I have a clear understanding of my breast cancer risk classification from my report | 394 (8.7%) | Understanding of Breast Cancer Report | Likert-Scale (5 options)<br><br>(`1`=Strongly Agree; `2`=Agree, `3`=Neither Agree nor Disagree, `4`=Disagree,`5`=Strongly Disagree) |
| I have a clear understanding of BREATHE study recommendations from my report. | 394 (8.7%) | Understanding of Breast Cancer Report | |
| I am confident that my breast cancer risk classification in my report is reliable. | 394 (8.7%) | Understanding of Breast Cancer Report | |
| Since receiving my report on breast cancer risk classification, I feel Relieved | 394 (8.7%) | Positive Emotional Reactions | |
| Since receiving my report on breast cancer risk classification, I feel Happy. | 394 (8.7%) | Positive Emotional Reactions | |
| Since receiving my report on breast cancer risk classification, I feel Motivated. | 394 (8.7%) | Positive Emotional Reactions | |
| Since receiving my report on breast cancer risk classification, I feel Regretful. | 394 (8.7%) | Negative Emotional Reactions | |
| Since receiving my report on breast cancer risk classification, I feel Disbelief. | 394 (8.7%) | Negative Emotional Reactions | |
| Since receiving my report on breast cancer risk classification, I feel Anxious. | 394 (8.7%) | Negative Emotional Reactions | |
| Since receiving my report on breast cancer risk classification, I feel Worried. | 394 (8.7%) | Negative Emotional Reactions | |
| Since receiving my report on breast cancer risk classification, I feel Stressed out. | 394 (8.7%) | Negative Emotional Reactions | |
| After learning about my breast cancer risk classification: What do you think is your chance of getting breast cancer? | 394 (8.7%) | *Perceived Breast Cancer Risk after Intervention* | Likert-Scale (7 options)<br><br>(1 = Strongly disagree, 2 = Disagree, 3 = Somewhat disagree, 4 = Neither |
| | | | agree nor disagree, 5 = Somewhat agree, 6 = Agree, 7 = Strongly agree) |
| Learning about my breast cancer risk classification has affected my ability to go on with my day-to-day task. | 394 (8.7%) | Behavioral Intentions after Intervention | Likert-Scale (5 options)<br>(`1`=Strongly Agree; `2`=Agree, `3`=Neither Agree nor Disagree, `4`=Disagree, `5`=Strongly Disagree) |
| After learning about my breast cancer risk classification: I will make changes to my lifestyle. | 394 (8.7%) | Behavioral Intentions after Intervention | |
| After learning about my breast cancer risk classification: I will make changes to my screening habits. | 394 (8.7%) | Behavioral Intentions after Intervention | |
| Knowing my risk classification including my genetic risk for developing cancer is important. | 394 (8.7%) | Attitudes towards genetics | |
| Knowing my risk classification including my genetic risk for cancer will motivate me to attend cancer screening according to my risk level | 394 (8.7%) | Attitudes towards genetics | |
| I would like to know my genetic risk classification for other health conditions, if available. | 394 (8.7%) | Attitudes towards genetics | |

Multimedia Appendix 2

**Table S1.** Pre-Intervention Survey Distributional Accuracy (Gemma 4 E4B)

| *Method*[a,b] | TVD | JSD | Time Taken [c] (s) |
|---|---|---|---|
| Random Baseline | 0.359 ± 0.016 | 0.361 ± 0.013 | - |
| Majority Vote Baseline | 0.238 | 0.322 | - |
| Zero Shot | 0.317 ± 0.012 | 0.318 ± 0.0097 | 4975 ± 45 |
| Age-Profile Agents | 0.237 ± 0.0067 | 0.250 ± 0.0068 | 5264 ± 66 |
| Minimal Profile Agents | 0.179 ± 0.0055 | 0.198 ± 0.0049 | 4888 ± 50 |
| Aggregate Profile Agents | 0.161 ± 0.0024 | 0.177 ± 0.0022 | 4925 ± 79 |
| Individual Profile Agents | 0.157 ± 0.0035 | 0.174 ± 0.0037 | 4583 ± 37 |
| Variation from Sampling n=50 | 0.052±0.012 | 0.071±0.004 | - |
| Split Sample Empirical Marginal (50/50) | 0.010 ± 0.001 | 0.012 ± 0.002 | |

[a]All experiments used the same underlying language model (Qwen3.5 9B). 5 runs were used.

[b] Soft Probability Aggregation was used to aggregate results.

[c] Time Taken to run 1 run of the experiment. 5000s ≈ 1.39 hrs

NWD not provided as only 3 out of 16 questions were ordinal type questions, the rest were nominal-type questions

**Table S2.** Pre-Intervention Survey Distributional Accuracy (Qwen3.5 9B)

| *Method*[a,b] | TVD | JSD | Time Taken [c] (s) |
|---|---|---|---|
| Majority Vote Baseline | 0.378 | 0.417 | - |
| Random Baseline | 0.359 ± 0.016 | 0.361 ± 0.013 | - |
| Zero Shot | 0.1346.± 0.004 | 0.142 ± 0.0039 | 1497 ± 4 |
| Age-Profile Agents | 0.122 ± 0.0017 | 0.1335 ± 0.0021 | 1603±137 |
| Minimal Profile Agents | 0.123 ± 0.0057 | 0.1332 ± 0.0047 | 1767± 170 |
| Aggregate Profile Agents | 0.1423 ± 0.0029 | 0.154 ± 0.0035 | 2157 ± 65 |
| Individual Profile Agents | 0.1368 ± 0.0061 | 0.145 ± 0.0067 | 2623 ± 318 |
| Variation from Sampling n=50 | 0.052±0.012 | 0.071±0.004 | - |
| Split Sample Empirical Marginal (50/50) | 0.010 ± 0.001 | 0.012 ± 0.002 | - |

[a]All experiments used the same underlying language model (Qwen3.5 9B). 5 runs were used.

[b] Soft Probability Aggregation was used to aggregate results.

[c] Time Taken to run 1 run of the experiment. 2000s ≈ 0.56 hrs

NWD not provided as only 3 out of 16 questions were ordinal type questions, the rest were nominal-type questions

**Table S3.** Semantic (Cosine) Similarity of reasoning by Agent Grounding Methods (Qwen3.5 9B)

| Variable | β | SE | 95% CI Lower | 95% CI Upper | p-value* |
|---|---|---|---|---|---|
| **Intercept** | 0.672 | 0.004 | 0.663 | 0.681 | <0.001 |
| **Age only Profile** | 0.015 | 0.004 | 0.008 | 0.023 | <0.001 |
| **Minimal Profile Agents** | 0.038 | 0.004 | 0.03 | 0.045 | <0.001 |
| **Aggregate Profile Agents** | 0.04 | 0.004 | 0.033 | 0.047 | <0.001 |
| **Individual Profile Agents** | 0.034 | 0.004 | 0.027 | 0.041 | <0.001 |
| **Run Variable** | 0.001 | 0.008 | - | - | - |

**Linear mixed-effects model with Question and run included as a random effect to account for repeated measures across grounding methods. Pairwise Cosine Similarity was averaged on a per-question, per-grounding method basis, before being given to the Linear mixed model.*

Zero-Shot Agents were used as the reference group.

**Linear Mixed Model Equation (For Cosine Similarity):**

$$C_{d,q} = \beta_0 + \beta_d + \mu_q + \epsilon_{d,q}$$

Where:

- $C$ is the cosine similarity
- $\beta_0$ is the fixed intercept term; representing the cosine similarity for the reference (zero-shot agents)
- $\beta_d$ is the estimated difference in cosine similarity between grounding method d and reference group (zero-shot agents)
- $\mu_q$ is the question-specific random intercept, where $\mu_q \sim N(0, \sigma_q^2)$
- $\epsilon_{d,q}$ is the residual error for grounding method ***d*** on question ***q***, where $\epsilon_{d,q} \sim N(0, \sigma^2)$

**Table S4.** Semantic (Cosine) Similarity of reasoning by Agent Grounding Methods (Gemma 4 E4B)

| Variable | β | SE | 95% CI Lower | 95% CI Upper | p-value* |
|---|---|---|---|---|---|
| **Intercept** | 0.598 | 0.004 | 0.589 | 0.606 | <0.001 |
| **Age only Profile** | 0.054 | 0.004 | 0.045 | 0.062 | <0.001 |
| **Minimal Profile Agents** | 0.058 | 0.004 | 0.05 | 0.067 | <0.001 |
| **Aggregate Profile Agents** | 0.077 | 0.004 | 0.069 | 0.085 | <0.001 |
| **Individual Profile Agents** | 0.068 | 0.004 | 0.059 | 0.076 | <0.001 |
| **Run Variable** | 0.001 | 0.005 | - | - | - |

* *Linear mixed-effects model with Question included as a random effect to account for repeated measures across grounding methods. Pairwise Cosine Similarity was averaged on a per-question, per-grounding method basis, before being given to the Linear mixed model.*

Zero-Shot Agents were used as the reference group.

**Linear Mixed Model Equation (For Cosine Similarity):**

$$\boldsymbol{C_{d,q} = \beta_0 + \beta_d + \mu_q + \epsilon_{d,q}}$$

Where:

- $C$ is the cosine similarity
- $\beta_0$ is the fixed intercept term; representing the cosine similarity for the reference (zero-shot agents)
- $\beta_d$ is the estimated difference in cosine similarity between grounding method d and reference group (zero-shot agents)
- $\mu_q$ is the question-specific random intercept, where $\mu_q \sim N(0, \sigma_q^2)$
- $\epsilon_{d,q}$ is the residual error for grounding method ***d*** on question ***q***, where $\epsilon_{d,q} \sim N(0, \sigma^2)$

**Table S5**. Post-Intervention Survey Distributional Accuracy (Gemma 4 E4B)

| *Method*[a,b] | TVD | JSD | NWD | Time Taken [c] (s) | % decrease in NWD relative to random | % decrease in NWD relative to zero-shot |
|---|---|---|---|---|---|---|
| Random Baseline | 0.409 ± 0.014 | 0.442 ± 0.01 | 0.230 ± 0.012 | - | 0.0 | |
| Majority Vote Baseline | 0.378 | 0.417 | 0.216 | - | -6.1 | |
| Zero Shot [d] | 0.487±0.02 | 0.461±0.02 | 0.222±0.005 | 7801±35 | -3.5 | 0.00 |
| Age-only Agents | 0.243±0.01 | 0.267±0.01 | 0.122±0.005 | 9960±68 | -47.0 | -45.05 |
| Minimal Profile Agents | 0.249±0.01 | 0.270±0.01 | 0.124±0.01 | 9788±124 | -46.1 | -44.14 |
| Aggregate Profile Agents | 0.1951±0.012 | 0.218±0.009 | 0.0928±0.0057 | 11500±73 | -59.7 | -58.20 |
| Individual Profile Agents | 0.1965±0.007 | 0.218±0.007 | 0.092±0.0048 | 11468±74 | -60.0 | -58.56 |
| Variation from Sampling n=50 | 0.106±0.01 | 0.134±0.01 | 0.037±0.004 | - | - | - |
| Split Sample Empirical Marginal (50/50) | 0.022±0.005 | 0.028 ± 0.004 | 0.008 ± 0.002 | - | - | - |

*For all metrics, lower is better. SD is calculated from 5 runs*

[a] All experiments used the same underlying language model (Gemma 4 E4B). 5 runs were used.

[b] Soft Probability Aggregation was used to aggregate results.

[c] Time Taken to run 1 run of the experiment. 10000s ≈ 2.78 hrs

[d] Three of the runs failed as the Gemma model refused to answer due to insufficient information.

**Table S6.** Post-Intervention Survey Distributional Accuracy (Qwen 3.5 9B)

| *Method*[a,b] | TVD | JSD | NWD | Time Taken [c] (s) | % decrease in NWD relative to random | % decrease in NWD relative to zero-shot |
|---|---|---|---|---|---|---|
| Random Baseline | 0.409 ± 0.014 | 0.442 ± 0.01 | 0.230 ± 0.012 | - | 0.0 | |
| Majority Vote Baseline | 0.378 | 0.417 | 0.216 | - | -6.1 | |
| Zero Shot | 0.396±0.0046 | 0.387±0.004 | 0.181±0.003 | 3026±18 | -21.3 | 0.00 |
| Age-only Agents | 0.285±0.016 | 0.298±0.015 | 0.127±0.0097 | 4449±75 | -44.8 | -29.83 |
| Minimal Profile Agents | 0.277±0.0075 | 0.290±0.0077 | 0.121±0.0052 | 4568±16 | -47.4 | -33.15 |
| Aggregate Profile Agents | 0.276±0.004 | 0.287±0.004 | 0.118±0.003 | 6240±39 | -48.7 | -34.81 |
| Individual Profile Agents | 0.252±0.0064 | 0.272±0.0073 | 0.110±0.006 | 6114±95 | -52.2 | -39.23 |
| Variation from Sampling n=50 | 0.106±0.01 | 0.134±0.01 | 0.037±0.004 | - | - | - |
| Split Sample Empirical Marginal (50/50) | 0.022±0.005 | 0.028 ± 0.004 | 0.008 ± 0.002 | - | - | - |

*For all metrics, lower is better. SD is calculated from 5 runs*

[a] All experiments used the same underlying language model (Qwen 3.5 9B). 5 runs were used.

[b] Soft Probability Aggregation was used to aggregate results.

[c] Time Taken to run 1 run of the experiment. 10000s ≈ 2.78 hrs

**Table S7.** Qwen3.5 9B Subgroup Difference. We include the non-pooled version here - we create a separate linear mixed model for subgroup analysis for each LLM.

| Category | Subgroup | Agents per run, n [a] | Mean NWD (SD) | β Coefficient[b] | Estimate [b] (95% CI) | Uncorrected P value [c] | Corrected P value [d] |
|---|---|---|---|---|---|---|---|
| **Age Group** | 35-39 | 22-26 | 0.109 (0.041) | Reference | Reference | - | - |
| | 40-44 | 44-47 | 0.116 (0.036) | 0.0065 | 0.01 (-0.01 to 0.02) | 0.302 | 1 |
| | 45-49 | 42-47 | 0.112 (0.034) | 0.0027 | 0.00 (-0.01 to 0.02) | 0.669 | 1 |
| | 50-54 | 40-52 | 0.114 (0.034) | 0.0042 | 0.00 (-0.01 to 0.02) | 0.507 | 1 |
| | 55-59 | 37-43 | 0.130 (0.035) | 0.0204 | 0.02 (0.01 to 0.03) | 0.001 | 0.015 |
| **Employed** | Yes | 163-177 | 0.113 (0.032) | Reference | Reference | - | - |
| | Never employed | 2-6 | 0.178 (0.090) | 0.0648 | 0.06 (0.04 to 0.09) | <0.001 | <0.001 |
| | Previously employed/retired | 21-31 | 0.107 (0.037) | -0.0058 | -0.01 (-0.03 to 0.02) | 0.508 | 1 |
| **Ethnicity** | Chinese | 153-154 | 0.115 (0.031) | Reference | Reference | - | - |
| | Indian | 11–20 | 0.116 (0.036) | 0.0012 | 0.00 (-0.01 to 0.02) | 0.876 | 1 |
| | Malay | 21–29 | 0.109 (0.039) | -0.0057 | -0.01 (-0.02 to 0.01) | 0.456 | 1 |
| | Others | 6–6 | 0.130 (0.051) | 0.015 | 0.01 (-0.00 to 0.03) | 0.051 | 0.514 |
| **Housing Type** | HDB 1-4 room | 81-88 | 0.108 (0.031) | Reference | Reference | - | - |
| | HDB Executive or 5-room flat | 67–73 | 0.113 (0.035) | 0.0056 | 0.01 (-0.00 to 0.02) | 0.253 | 1 |
| | Private Property and Others | 45–46 | 0.122 (0.038) | 0.0141 | 0.01 (0.00 to 0.02) | 0.004 | 0.044 |
| **Marital Status** | Currently Married | 153-162 | 0.111 (0.032) | Reference | Reference | - | - |
| | Never Married | 21–29 | 0.121 (0.040) | 0.0101 | 0.01 (-0.01 to 0.03) | 0.26 | 1 |
| | Divorced | 16–17 | 0.114 (0.043) | 0.0029 | 0.00 (-0.01 to 0.02) | 0.749 | 1 |
| | Widowed | 2–2 | 0.208 (0.088) | 0.0996 | 0.10 (0.08 to 0.12) | <0.001 | <0.001 |

**[a]** Agents per run reflects the pooled sample used in the mixed-effects model. Unlike previous analyses where only n=50 agents were used, 200 agents were generated for each LLM across 2 runs

[b] Estimated change in NWD relative to the reference subgroup.

[c] Linear Mixed Effects Model (Multimedia Appendix 3) included demographic category as a fixed effect; 'question' (question number) and 'run' (run 1, run 2) were included as random effects to account for repeated questions and run to run variability. Individual profile agents were used as it allows comparison of subgroup-specific NWDs with the corresponding subgroup ground-truth distributions.

**[d]** Adjusted via Holm-Bonferroni Correction. To control for family-wise error rates across multiple demographic categories representing multiple hypothesis, raw p-values from reference group comparisons across all demographic variables were pooled into a single vector and adjusted.

**Table S8.** Gemma 4 E4B Subgroup Difference. We include the non-pooled version here - we create a separate linear mixed model for subgroup analysis for each LLM.

| Category | Subgroup | Agents per run, n [a] | Mean NWD (SD) | β Coefficient[b] | Estimate [b] (95% CI) | Uncorrected P value [c] | Corrected P value [d] |
|---|---|---|---|---|---|---|---|
| **Age Group** | 35-39 | 22-26 | 0.086 (0.036) | Reference | Reference | - | - |
| | 40-44 | 44-47 | 0.088 (0.040) | 0.0019 | 0.00 (-0.01 to 0.02) | 0.804 | 1 |
| | 45-49 | 42-47 | 0.084 (0.038) | -0.0015 | -0.00 (-0.02 to 0.01) | 0.845 | 1 |
| | 50-54 | 40-52 | 0.103 (0.059) | 0.0171 | 0.02 (0.00 to 0.03) | 0.023 | 0.231 |
| | 55-59 | 37-43 | 0.108 (0.059) | 0.0225 | 0.02 (0.01 to 0.04) | 0.003 | 0.033 |
| **Employed** | Yes | 163-177 | 0.088 (0.044) | Reference | Reference | - | - |
| | Never employed | 2-6 | 0.170 (0.13) | 0.082 | 0.08 (0.05 to 0.11) | <0.001 | <0.001 |
| | Previously employed/retired | 21-31 | 0.087 (0.037) | -0.00037 | -0.00 (-0.03 to 0.03) | 0.982 | 1 |
| **Ethnicity** | Chinese | 153-154 | 0.089 (0.045) | Reference | Reference | - | - |
| | Indian | 11–20 | 0.101 (0.057) | 0.0012 | 0.00 (-0.01 to 0.02) | 0.157 | 1 |
| | Malay | 21–29 | 0.085 (0.039) | -0.0057 | -0.01 (-0.02 to 0.01) | 0.662 | 1 |
| | Others | 6–6 | 0.109 (0.045) | 0.015 | 0.01 (-0.00 to 0.03) | 0.015 | 0.169 |
| **Housing Type** | HDB 1-4 room | 81-88 | 0.090 (0.041) | Reference | Reference | - | - |
| | HDB Executive or 5-room flat | 67–73 | 0.085 (0.042) | -0.0053 | -0.01 (-0.02 to 0.01) | 0.32 | 1 |
| | Private Property and Others | 45–46 | 0.098 (0.051) | 0.0081 | 0.01 (-0.00 to 0.02) | 0.131 | 1 |
| **Marital Status** | Currently Married | 153-162 | 0.087 (0.043) | Reference | Reference | - | - |
| | Never Married | 21–29 | 0.090 (0.035) | 0.0035 | 0.00 (-0.01 to 0.02) | 0.685 | 1 |
| | Divorced | 16–17 | 0.103 (0.054) | 0.0159 | 0.02 (-0.00 to 0.03) | 0.062 | 0.56 |
| | Widowed | 2–2 | 0.164 (0.081) | 0.0783 | 0.08 (0.06 to 0.10) | <0.001 | <0.001 |

[a] Agents per run reflects the pooled sample used in the mixed-effects model. Unlike previous analyses where only n=50 agents were used, 200 agents were generated for each LLM across 2 runs

[b] Estimated change in NWD relative to the reference subgroup.

[c] Linear Mixed Effects Model (Multimedia Appendix 3) included demographic category as a fixed effect; 'question' (question number) and 'run' (run 1, run 2) were included as random effects to account for repeated questions and run to run variability. Individual profile agents were used as it allows comparison of subgroup-specific NWDs with the corresponding subgroup ground-truth distributions.

[d] Adjusted via Holm-Bonferroni Correction. To control for family-wise error rates across multiple demographic categories representing multiple hypothesis, raw p-values from reference group comparisons across all demographic variables were pooled into a single vector and adjusted.

**Table S9.** Probability Aggregation vs Single Choice Selection

**A: NWD**

| *Model Used* | *Method* | NWD (Soft) | NWD (Hard) | Δ NWD (Δ %) |
|---|---|---|---|---|
| **Gemma 4 E4B** | Zero Shot | 0.222±0.005 | 0.376±0.003 | -0.154 (-41.1%) |
| | Age-only Agents | 0.122±0.005 | 0.135±0.006 | -0.013 (-9.7%) |
| | Minimal Profile Agents | 0.124±0.01 | 0.136±0.008 | -0.012 (-8.98%) |
| | Aggregate Profile Agents | 0.0928±0.0057 | 0.109±0.007 | -0.0162 (-14.6%) |
| | Individual Profile Agents | 0.092±0.0048 | 0.109±0.006 | -0.017 (-15.7%) |
| **Qwen 3.5-9B** | | | | |
| | Zero Shot | 0.181±0.003 | 0.188±0.006 | -0.007 (-3.88%) |
| | Age-only Agents | 0.127±0.0097 | 0.144±0.01 | -0.017 (-11.51%) |
| | Minimal Profile Agents | 0.121±0.0052 | 0.139±0.005 | -0.019 (-13.3%) |
| | Aggregate Profile Agents | 0.110±0.003 | 0.116±0.009 | -0.006 (-5.0%) |
| | Individual Profile Agents | 0.118±0.006 | 0.125±0.002 | -0.008 (-6.1%) |

**B: JSD**

| *Model Used* | *Method* | JSD (Soft) | JSD (Hard) |
|---|---|---|---|
| **Gemma 4 E4B** | | | |
| | Zero Shot | 0.461±0.02 | 0.765±0.004 |
| | Age-only Agents | 0.267±0.01 | 0.335±0.02 |
| | Minimal Profile Agents | 0.270±0.01 | 0.346±0.009 |
| | Aggregate Profile Agents | 0.218±0.009 | 0.2968±0.008 |
| | Individual Profile Agents | 0.218±0.007 | 0.2895±0.01 |
| **Qwen 3.5-9B** | | | |
| | Zero Shot | 0.387±0.004 | 0.489±0.008 |
| | Age-only Agents | 0.298±0.015 | 0.388±0.02 |
| | Minimal Profile Agents | 0.290±0.0077 | 0.389±0.013 |
| | Aggregate Profile Agents | 0.287±0.004 | 0.365±0.007 |
| | Individual Profile Agents | 0.272±0.0073 | 0.347±0.015 |

**C: TVD**

| *Model Used* | *Method* | TVD (Soft) | TVD (Hard) |
|---|---|---|---|
| **Gemma 4 E4B** | | | |
| | Zero Shot | 0.487±0.02 | 0.758±0.006 |
| | Age-only Agents | 0.243±0.01 | 0.300±0.021 |
| | Minimal Profile Agents | 0.249±0.01 | 0.317±0.012 |
| | Aggregate Profile Agents | 0.1951±0.012 | 0.258±0.011 |
| | Individual Profile Agents | 0.1965±0.007 | 0.243±0.013 |
| **Qwen 3.5-9B** | | | |
| | Zero Shot | 0.396±0.0046 | 0.486±0.011 |
| | Age-only Agents | 0.285±0.016 | 0.376±0.027 |
| | Minimal Profile Agents | 0.277±0.0075 | 0.379±0.017 |
| | Aggregate Profile Agents | 0.276±0.004 | 0.368±0.009 |
| | Individual Profile Agents | 0.252±0.0064 | 0.329±0.015 |

All results were computed on the post-intervention questionnaire

**Table S10.** Prompt Sensitivity Results

| *Metric* | *Method* [a,b] | Prompt Template (Variant 1) | Prompt Template (Variant 2) | Mean Δ (SD) | % Change (SD) |
|---|---|---|---|---|---|
| *TVD* | *Individual Profile* | 0.2533 | 0.2504 | -0.0029 (0.0150) | -1.0 (5.7) % |
| | *Aggregate Profile* | 0.2756 | 0.2712 | -0.0044 (0.0092) | -1.5 (3.3) % |
| | *Age-Only Profile* | 0.2851 | 0.2758 | -0.0093 (0.0149) | -3.0 (5.3) % |
| | *Minimal Profile* | 0.2763 | 0.2653 | -0.0010 (0.0149) | -3.9 (2.7) % |
| | *Zero-Shot* | 0.3956 | 0.3803 | -0.0153 (0.0069) | -3.9 (1.8) % |
| *JSD* | *Individual Profile* | 0.2726 | 0.2655 | -0.0071 (0.0153) | -2.5 (5.3) % |
| | *Aggregate Profile* | 0.2874 | 0.2788 | -0.0087 (0.0118) | -2.95 (4.05) % |
| | *Age-Only Profile* | 0.2976 | 0.2804 | -0.0172 (0.0136) | -5.6 (4.4) % |
| | *Minimal Profile* | 0.2902 | 0.2723 | -0.0178 (0.0073) | -6.1 (2.4) % |
| | *Zero-Shot* | 0.3867 | 0.3721 | -0.0146 (0.0084) | -3.8 (2.2) % |
| *NWD* | *Individual Profile* | 0.1114 | 0.1197 | 0.0083 (0.0089) | 7.8 (8.3) % |
| | *Aggregate Profile* | 0.1176 | 0.1233 | 0.0057 (0.0071) | 4.9 (6.1) % |
| | *Age-Only Profile* | 0.1273 | 0.1221 | -0.0051 (0.0090) | -3.6 (7.2) % |
| | *Minimal Profile* | 0.1207 | 0.1158 | -0.0048 (0.0047) | -3.9 (3.7) % |
| | *Zero-Shot* | 0.1816 | 0.1743 | -0.0073 (0.0059) | -4.0 (3.2) % |

*For all metrics, lower is better.*

[a] All experiments used the same underlying language model (Qwen 3.5 9B) on the post-intervention survey. 5 runs were used.

[b] Soft Probability Aggregation was used to aggregate results.

**Table S11.** Temperature Sensitivity Results

| *Metric* | *Method* [a,b] | Mean Metric for Temperature 0.6 | Mean Metric for Temperature 0 | Mean Δ (SD) | % Change (SD) |
|---|---|---|---|---|---|
| *TVD* | *Individual Profile* | 0.2504 | 0.3162 | 0.066 (0.016) | 20.7 (4.5) % |
| | *Aggregate Profile* | 0.2712 | 0.2968 | 0.026 (0.005) | 8.6 (1.5) % |
| | *Age-Only Profile* | 0.2758 | 0.3078 | 0.032 (0.004) | 10.3 (1.1) % |
| | *Minimal Profile* | 0.2653 | 0.2792 | 0.014 (0.004) | 4.9 (1.6) % |
| | *Zero-Shot* | 0.3803 | 0.4567 | 0.076 (0.008) | 16.7 (1.9) % |
| *JSD* | *Individual Profile* | 0.2655 | 0.3097 | 0.044 (0.013) | 14.2 (4.0) % |
| | *Aggregate Profile* | 0.2788 | 0.2922 | 0.013 (0.006) | 4.6 (2.1) % |
| | *Age-Only Profile* | 0.2804 | 0.3074 | 0.027 (0.003) | 8.8 (0.9) % |
| | *Minimal Profile* | 0.2723 | 0.2867 | 0.014 (0.004) | 5.0 (1.3) % |
| | *Zero-Shot* | 0.3721 | 0.4474 | 0.075 (0.007) | 16.8 (1.7) % |
| *NWD* | *Individual Profile* | 0.1197 | 0.1339 | 0.014 (0.007) | 10.5 (5.0) % |
| | *Aggregate Profile* | 0.1233 | 0.1194 | -0.004 (0.004) | -3.3 (3.2) % |
| | *Age-Only Profile* | 0.1221 | 0.1359 | 0.014 (0.002) | 10.1 (1.1) % |
| | *Minimal Profile* | 0.1158 | 0.1202 | 0.004 (0.002) | 3.6 (1.9) % |
| | *Zero-Shot* | 0.1743 | 0.2215 | 0.047 (0.005) | 21.3 (2.2) % |

*For all metrics, lower is better.*

[a] All experiments used the same underlying language model (Qwen 3.5 9B) on the post-intervention survey with Prompt Template Variant 2. 5 runs were used.

[b] Soft Probability Aggregation was used to aggregate results.

**Table S12.** Reasoning Length by Agent Grounding Methods (Qwen3.5 9B)

| Variable | $\hat{\beta}$ | SE | 95% CI Lower | 95% CI Upper | p-value |
|---|---|---|---|---|---|
| **Intercept** | 55.9 | 0.78 | 54.4 | 57.44 | <0.001 |
| **Age only Profile** | 11.29 | 0.99 | 9.35 | 13.23 | <0.001 |
| **Minimal Profile Agents** | 21.09 | 0.99 | 19.15 | 23.03 | <0.001 |
| **Aggregate Profile Agents** | 28.57 | 0.99 | 26.63 | 30.5 | <0.001 |
| **Individual Profile Agents** | 37.36 | 0.99 | 35.36 | 39.23 | <0.001 |
| **Run Variable** | 12.5 | 0.51 | - | - | - |

**Linear mixed-effects model with Question included as a random effect.*

*$\hat{\beta}$ represents estimated differences in the actual reasoning word length (words) contained in the JSON reasoning output, rather than the total number of tokens generated. Zero-Shot Agents served as the reference group.*

**Linear Mixed Model Equation (For Reasoning Length):**

$$R_{d,q} = \beta_0 + \beta_d + \mu_q + \epsilon_{d,q}$$

Where:

- $R$ is the average number of words
- $\beta_0$ is the fixed intercept term; representing the average number of words for the reference (zero-shot agents)
- $\beta_d$ is the estimated difference in number of words between grounding method d and reference group (zero-shot agents)
- $\mu_q$ is the question-specific random intercept, where $\mu_q \sim N(0, \sigma_q^2)$
- $\epsilon_{d,q}$ is the residual error for grounding method ***d*** on question ***q***, where $\epsilon_{d,q} \sim N(0, \sigma^2)$

**Table S13.** Reasoning Length by Agent Grounding Methods (Gemma 4 E4B)

| Variable | β̂ | SE | 95% CI Lower | 95% CI Upper | p-value |
|---|---|---|---|---|---|
| **Intercept** | 49.16 | 0.56 | 48.7 | 50.3 | <0.001 |
| **Age only Profile** | 3.45 | 0.55 | 2.36 | 4.53 | <0.001 |
| **Minimal Profile Agents** | 4.95 | 0.55 | 3.86 | 6.03 | <0.001 |
| **Aggregate Profile Agents** | 7.84 | 0.55 | 6.75 | 8.92 | <0.001 |
| **Individual Profile Agents** | 8.92 | 0.55 | 7.83 | 10 | <0.001 |
| **Run Variable** | 8.71 | 0.6 | - | - | - |

**Linear mixed-effects model with Question included as a random effect.*

***β̂*** *represents estimated differences in the actual reasoning word length (words) contained in the JSON reasoning output, rather than the total number of tokens generated. Zero-Shot Agents served as the reference group.*

**Linear Mixed Model Equation (For Reasoning Length):**

$$R_{d,q} = \beta_0 + \beta_d + \mu_q + \epsilon_{d,q}$$

Where:

- $R$ is the average number of words
- $\beta_0$ is the fixed intercept term; representing the average number of words for the reference (zero-shot agents)
- $\beta_d$ is the estimated difference in number of words between grounding method d and reference group (zero-shot agents)
- $\mu_q$ is the question-specific random intercept, where $\mu_q \sim N(0, \sigma_q^2)$
- $\epsilon_{d,q}$ is the residual error for grounding method ***d*** on question ***q***, where $\epsilon_{d,q} \sim N(0, \sigma^2)$

**Figure S1.** Post-Intervention Responses Mean Likert Response (Qwen3.5 9B model). *Mean responses averaged over 5 runs. It is important to note that, except for the questions with the theme "general well-being" and "perceived breast cancer risk after intervention" which are on the 7-point Likert scale, the other questions are on the 5-point Likert scale.*

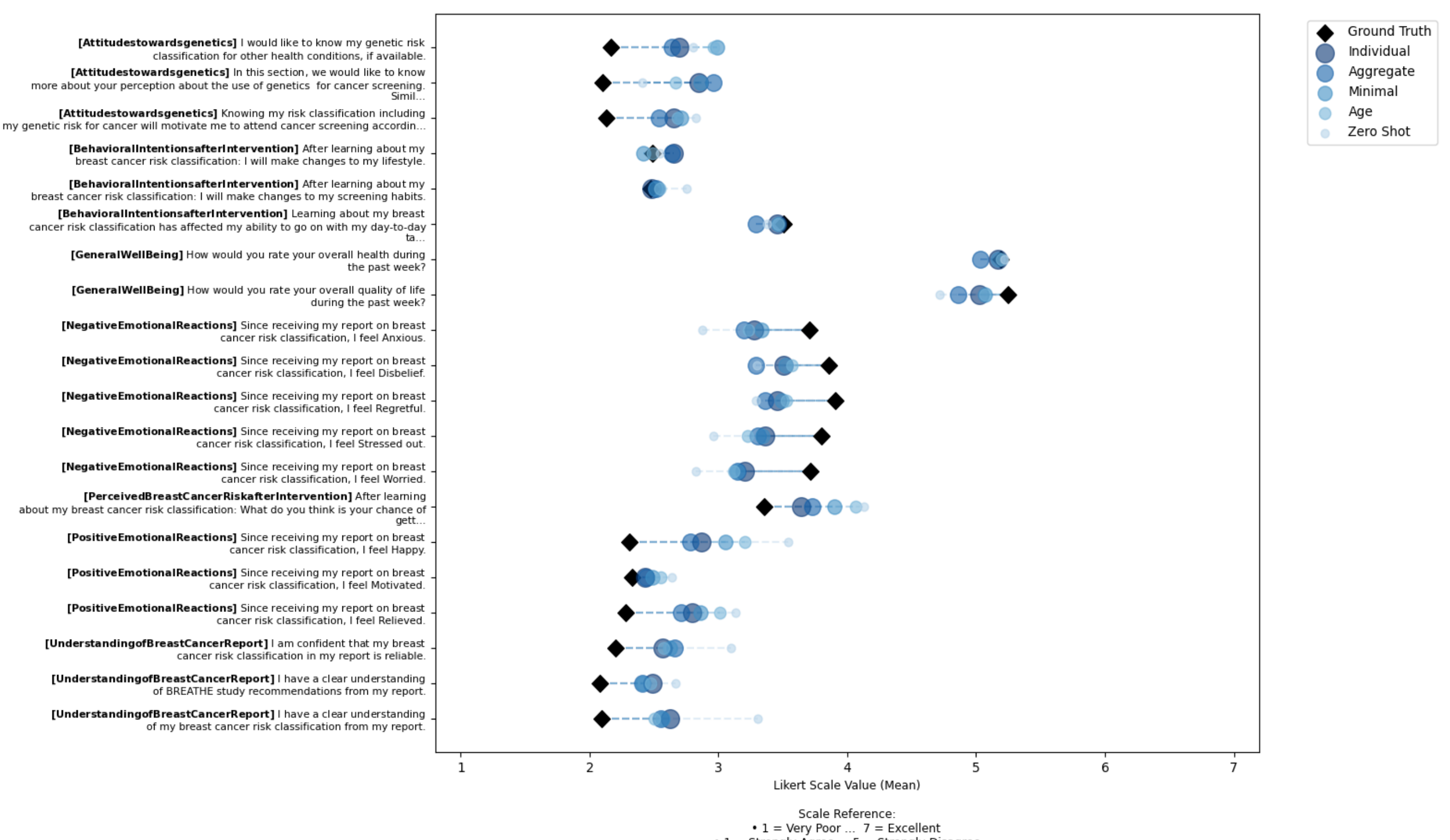

**Figure S2**. Post-Intervention Responses Mean Likert Response (Gemma 4 E4B model). *Mean responses averaged over 5 runs. It is important to note that, except for the questions with the theme "general well-being" and "perceived breast cancer risk after intervention" which are on the 7-point Likert scale, the other questions are on the 5-point Likert scale.*

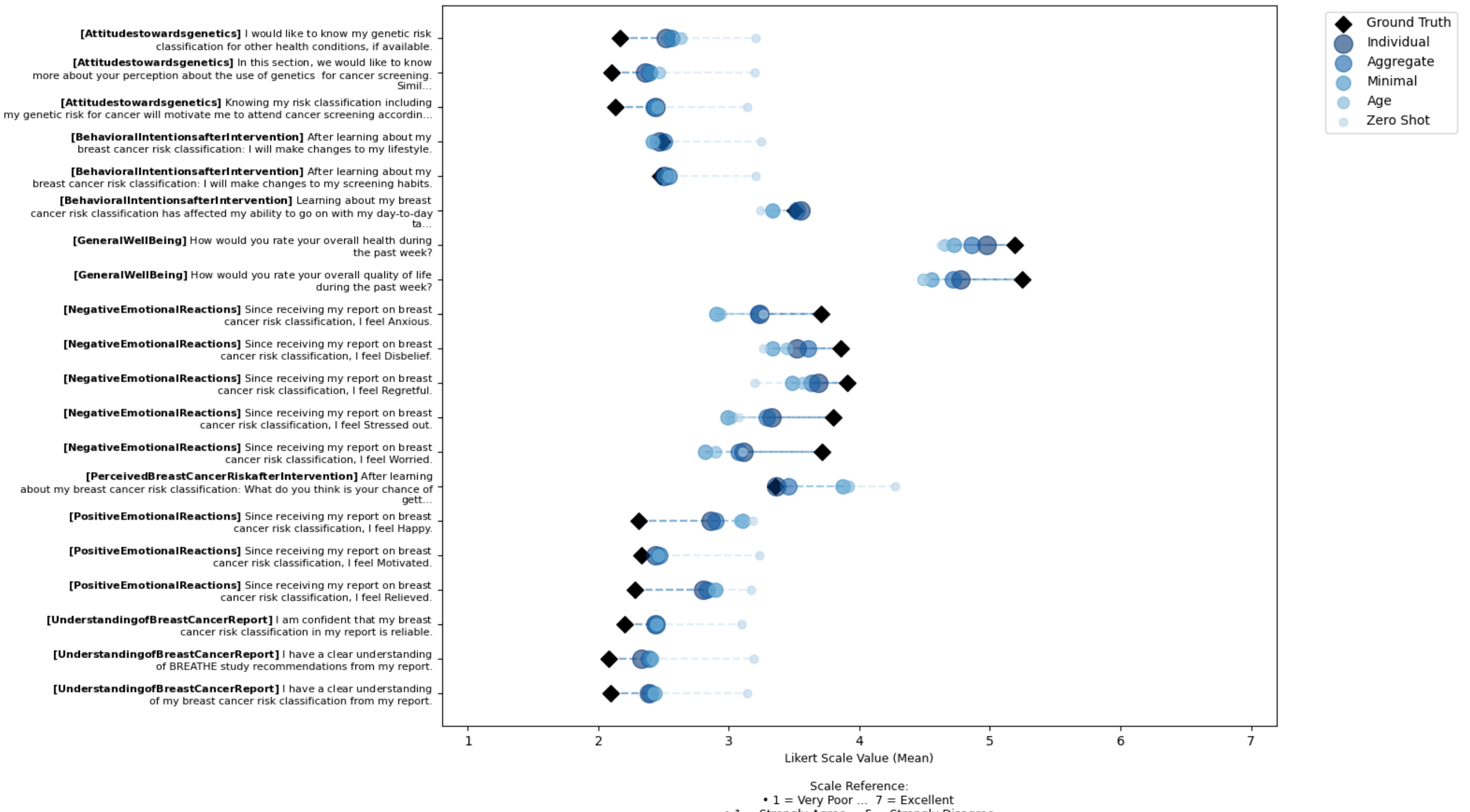

**Figure S3. Pre-Intervention** Prediction Errors by Question Theme (Qwen-3.5 9b)

| | TVD | JSD |
|---|---|---|
| Risk Perception | 0.089±0.008 | 0.112±0.011 |
| Prevention and Screening Knowledge | 0.122±0.02 | 0.141±0.021 |
| Screening Barriers | 0.124±0.013 | 0.122±0.016 |
| Cancer Fatalism | 0.187±0.018 | 0.187±0.015 |

**Figure S4.** Pre-Intervention Overall Responses (Qwen 3.5 9B model). *Mean Responses averaged over 5 runs.*

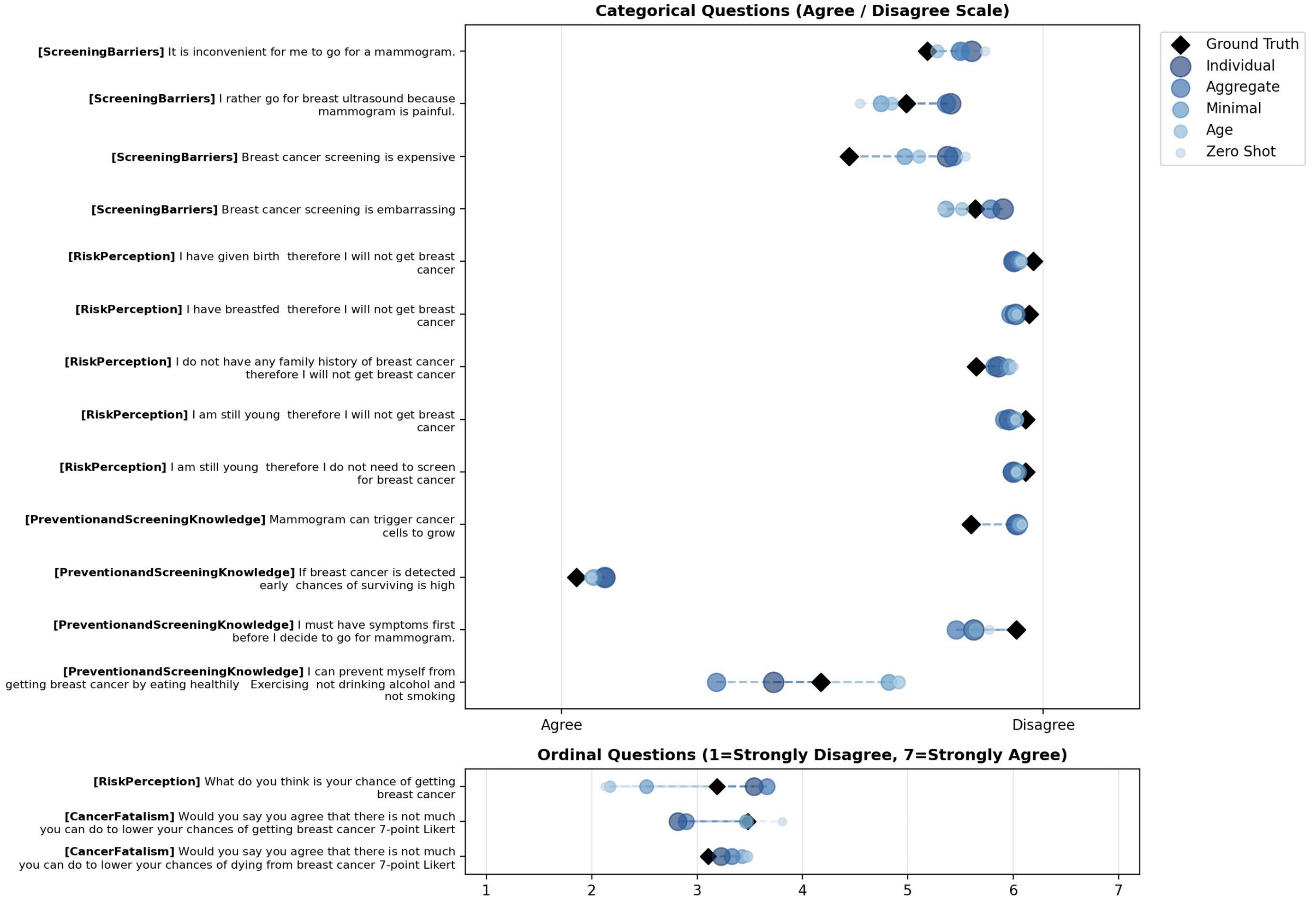

**Figure S5.** Pre-Intervention Overall Responses (Gemma 4 E4B model). *Mean Responses averaged over 5 runs.*

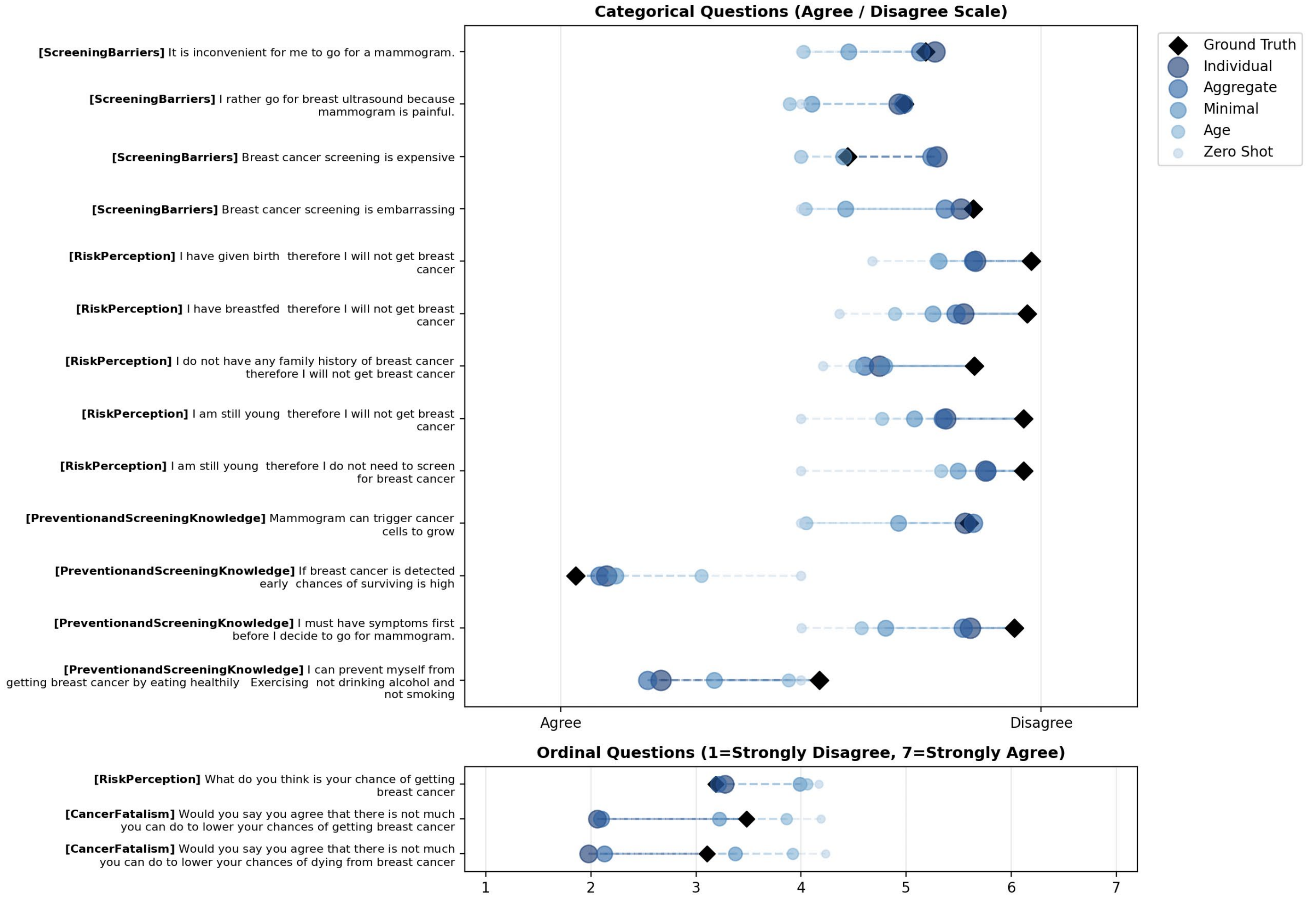

**Figure S6. Post-Intervention** Prediction Errors by Question Theme (Qwen 3.5 9B)

| | TVD | JSD | NWD |
|---|---|---|---|
| General Well Being | 0.169±0.007 | 0.19±0.009 | 0.058±0.008 |
| Perceived Breast Cancer Risk after Intervention | 0.245±0.028 | 0.267±0.029 | 0.071±0.011 |
| Positive Emotional Reactions | 0.254±0.017 | 0.277±0.017 | 0.109±0.012 |
| Behavioral Intentions after Intervention | 0.255±0.02 | 0.247±0.017 | 0.08±0.007 |
| Negative Emotional Reactions | 0.273±0.019 | 0.271±0.014 | 0.149±0.009 |
| Understanding of Breast Cancer Report | 0.323±0.015 | 0.345±0.014 | 0.123±0.007 |
| Attitudes towards genetics | 0.357±0.032 | 0.379±0.022 | 0.161±0.014 |

Relative Performance for Metric
(Red = Worse, Green = Better)

**Figure S7**. Effect of Number of Agents on Distributional Error

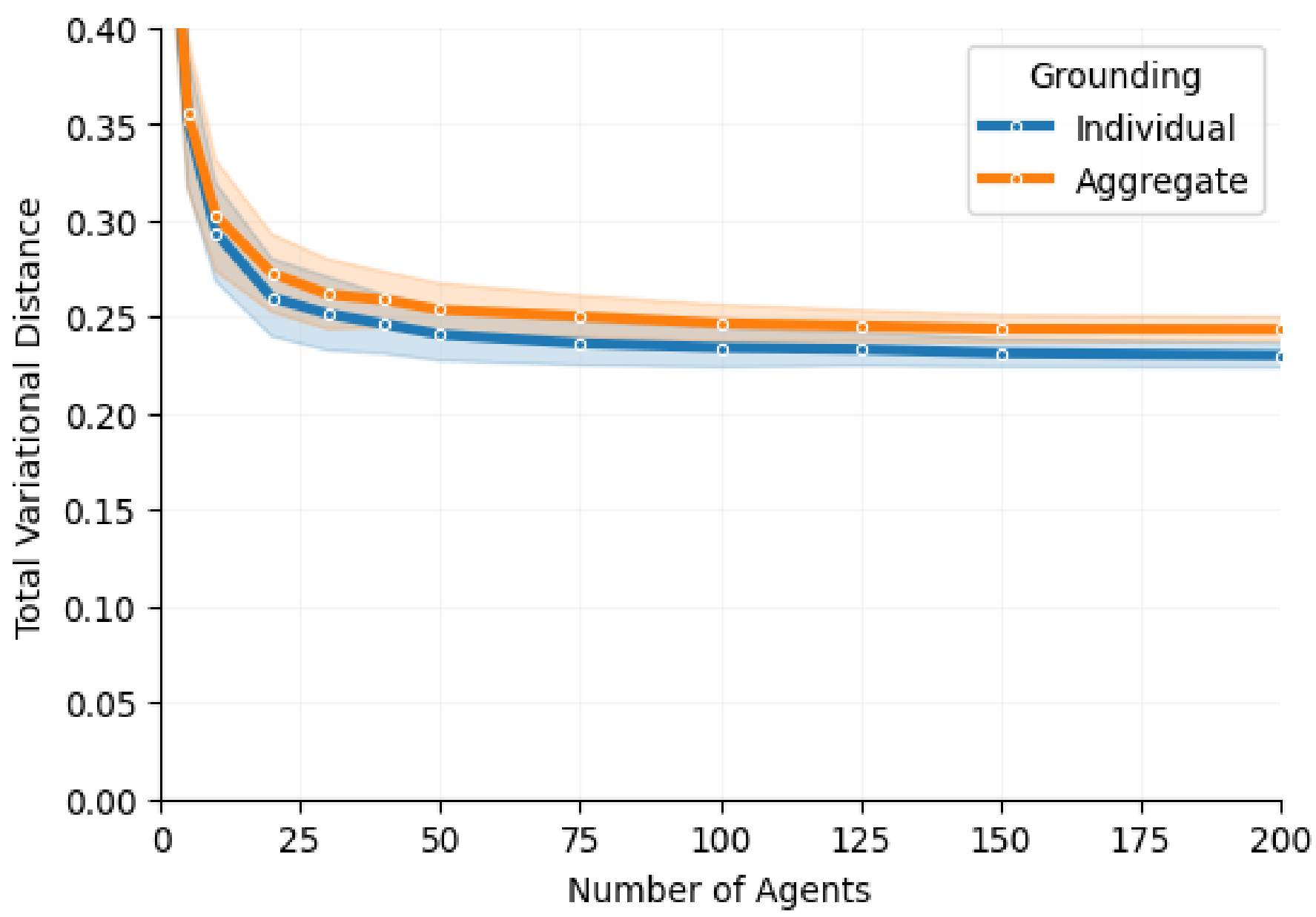

Shaded uncertainty region(s) represents the Standard Deviation. Computed from responses to post-intervention survey, which were aggregated by soft probability aggregation. 

For number of agents required, subsets of size ∈ {2,5,10,20,30,40,50,75,100,125,150,200} were drawn with replacement across 200 iterations from the pooled agent base (agent responses across 5 independent simulation runs), ensuring that there was a new cohort drawn for each subset. It is important to note that there are no constraints to ensure that the sample matches the original true distribution, other than the fact that the underlying pooled agent base generally matches the population for marginal distributions. This plot is consistent with our informal observations, though it is not a replacement for actual experimentation.

**Extra:** Radar Plots for pre and post intervention prediction errors by Question Theme

**Qwen3.5- 9B**

The plots shown here are based on similar results as the heatmaps presented earlier, just in a different style of visualization (only TVD is displayed here).

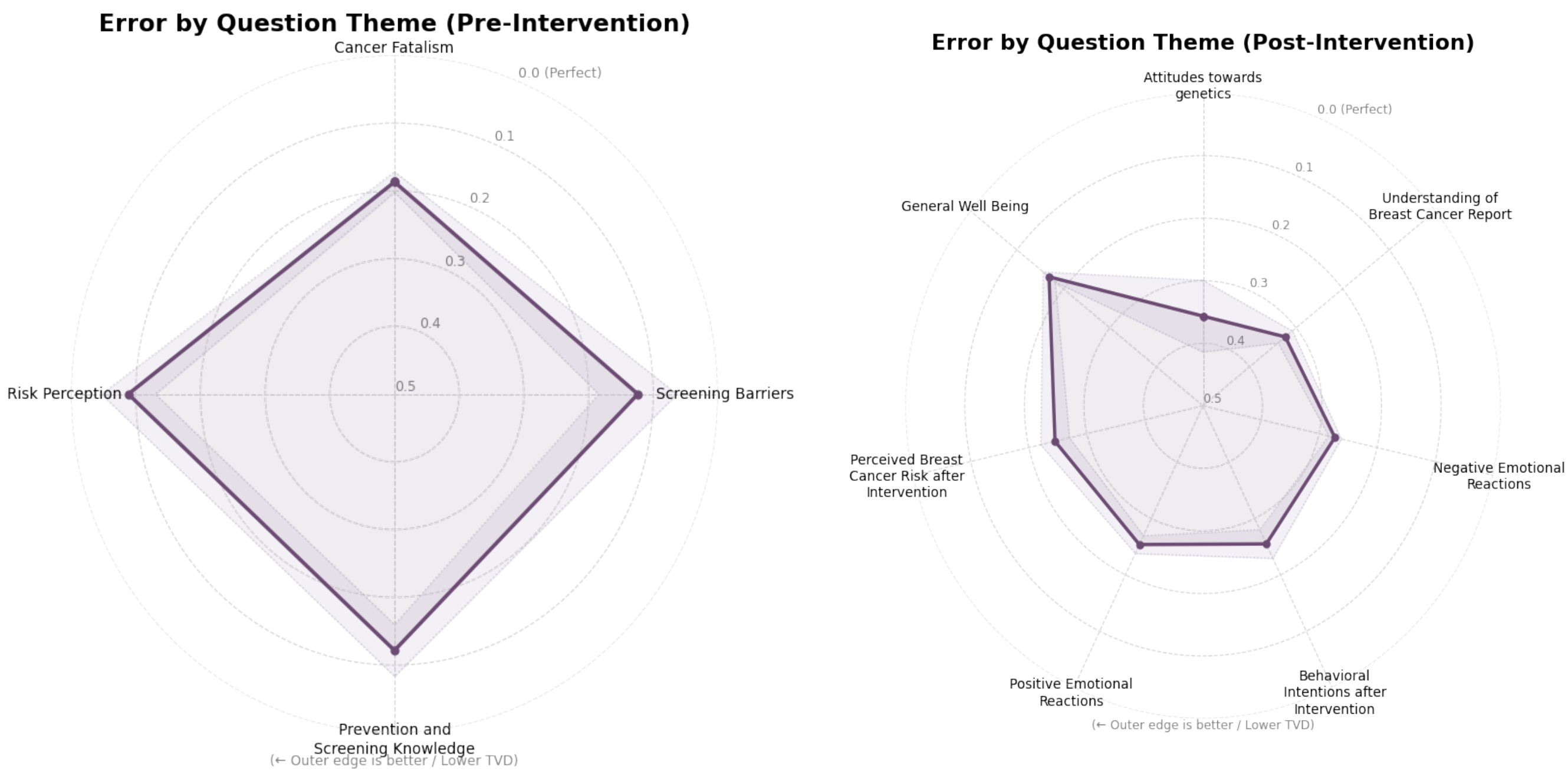

**Gemma 4 E4B**

The plots shown here are based on similar results as the heatmaps presented earlier, just in a different style of visualization (only TVD is displayed here).

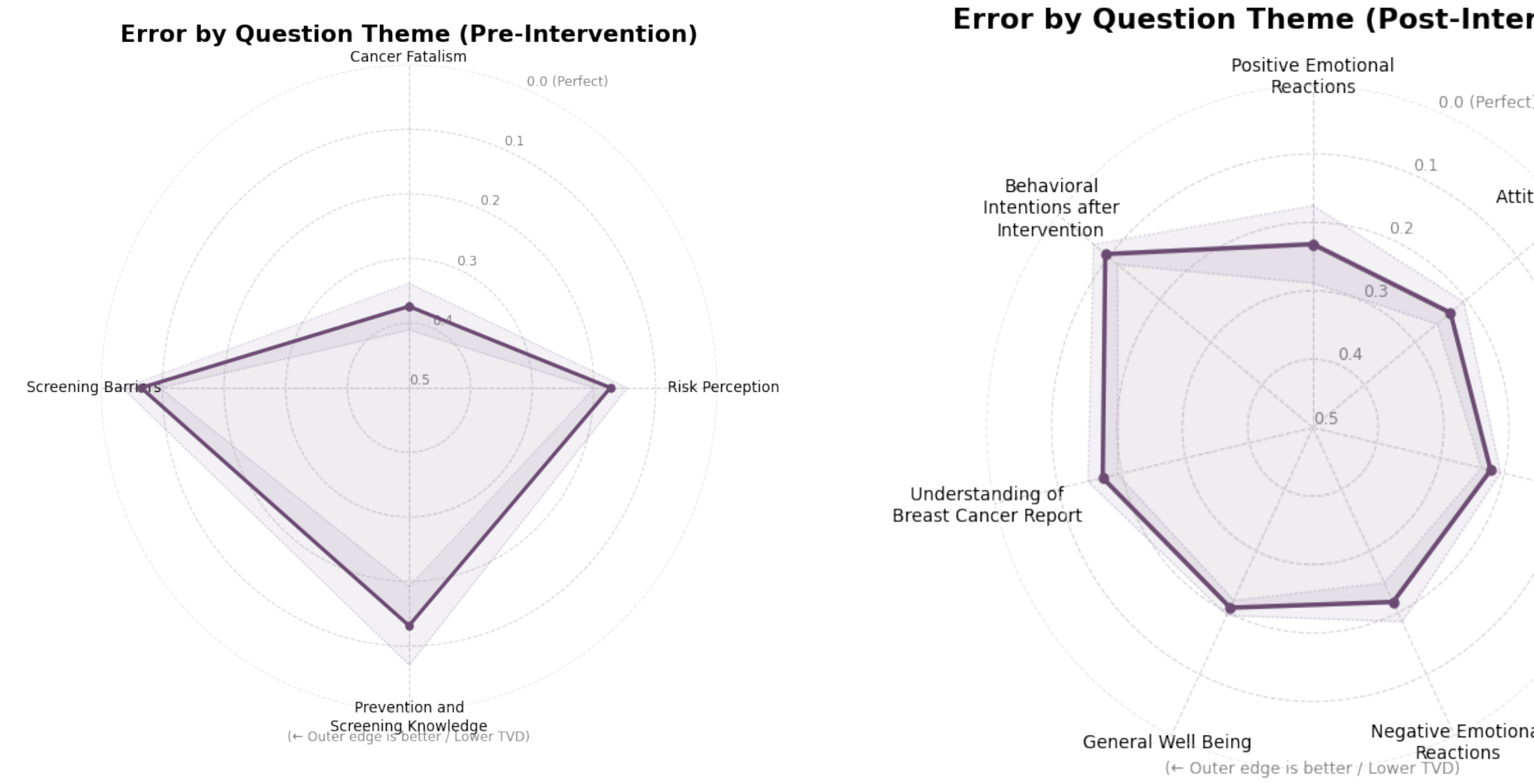

**Multimedia Appendix 3.**

## A. Distributional Similarity Metrics

*Total Variation Distance (TVD)* measures the overall discrepancy between two distributions.

$$TVD = \frac{1}{2} \sum_{x \in \Omega} |P(x) - Q(x)|$$

where $P$ and $Q$ refer to the predicted and actual response distributions (discrete probability distributions). TVD is bounded between 0 and 1 and scale-invariant regardless of the number of options, and thus applicable to the discrete response categories found in survey questions.

*Jensen Shannon Distance (JSD)* is a symmetric information theoretic measure of divergence between probability distributions. It is defined as:

$$JSD(P, Q) = \sqrt{\frac{1}{2} D_{KL}(P \parallel M) + \frac{1}{2} D_{KL}(Q \parallel M)}$$

Where

- $M = \frac{P+Q}{2}$
- KL Divergence (Discrete) is defined as: $D_{KL}(P \parallel M) = \sum_x P(x) \log_2 \frac{P(x)}{M(x)}$

Like TVD it is also scale invariant and bounded between 0 and 1. Thus JSD and TVD are complementary metrics of distributional differences.

*Normalised Wasserstein Distance (NWD)*. Unlike TVD and JSD, Wasserstein Distance penalizes distant errors more heavily than near misses. Nominal questions lack inherent ordering, thus NWD is only calculated for ordinal questions.

$$WD_{norm} = \frac{1}{x_k - x_1} \left[ \sum_{i=1}^{k-1} |F(x_i) - G(x_i)| \, (x_{i+1} - x_i) \right]$$

- $F$ and $G$ are cumulative distribution functions for the predicted and ground truth distributions respectively

As shown above, to enable comparison across questions with different number of categories (scale invariance), we normalize NWD ($WD_{norm}$) between 0 and 1 by dividing $WD$ by $x_k - x_1$ (equivalent to dividing by $k - 1$ categories). This normalization assumes the Likert categories are equally spaced throughout the range, which is a valid assumption for this application.

## B. Linear Mixed Model Equations

**Linear Mixed Model Equation (For Subgroups):**

$$\boldsymbol{NWD_{d,q,r} = \beta_0 + \beta_d + \mu_q + v_r + \epsilon_{d,q,r}}$$

where:

- $NWD_{d,q,r}$ is the NWD for subgroup d, question q, run r
- $\beta_0$ is the intercept term; representing the NWD for the reference subgroup
- $\beta_d$ is the fixed effect representing the mean difference in NWD between subgroup *d* and the reference subgroup.
- $\mu_q$ is a random variable, which is the random intercept for each question *q,* where $\mu_q \sim N(0, \sigma_q^2)$
- $v_r$ is a random variable, the random intercept for each run *r,* where $v_r \sim N(0, \sigma_r^2)$
- $\epsilon_{d,q,r}$ is the residual error where $\epsilon_{d,q,r} \sim N(0, \sigma^2)$ for question q, run r, subgroup d

Each demographic variable was fitted with a separate linear mixed model.

**Linear Mixed Model Equation (For Reasoning Length):**

$$\boldsymbol{R_{d,q} = \beta_0 + \beta_d + \mu_q + \epsilon_{d,q}}$$

Where:

- $R$ is the average number of words
- $\beta_0$ is the fixed intercept term; representing the average number of words for the reference (zero-shot agents)
- $\beta_d$ is the estimated difference in number of words between grounding method d and reference group (zero-shot agents)
- $\mu_q$ is the question-specific random intercept, where $\mu_q \sim N(0, \sigma_q^2)$
- $\epsilon_{d,q}$ is the residual error for grounding method *d* on question *q*, where $\epsilon_{d,q} \sim N(0, \sigma^2)$

**Linear Mixed Model Equation (For Cosine Similarity):**

$$\boldsymbol{C_{d,q} = \beta_0 + \beta_d + \mu_q + \epsilon_{d,q}}$$

Where:

- $C$ is the cosine similarity
- $\beta_0$ is the fixed intercept term; representing the cosine similarity for the reference (zero-shot agents)
- $\beta_d$ is the estimated difference in cosine similarity between grounding method d and reference group (zero-shot agents)
- $\mu_q$ is the question-specific random intercept, where $\mu_q \sim N(0, \sigma_q^2)$
- $\epsilon_{d,q}$ is the residual error for grounding method *d* on question *q*, where $\epsilon_{d,q} \sim N(0, \sigma^2)$

**Multimedia Appendix 4**

Table S1. Topic Modelling for the Question ‘Would you say you agree there is not much you can do to lower your chances of dying from breast cancer’

Table S2. Topic Modelling for the Question ‘Breast Cancer Screening is Expensive’

Table S3. Sample LLM Outputs

**Table S1.** Topic Modelling for the Question 'Would you say you agree there is not much you can do to lower your chances of dying from breast cancer'

| Topic | Count | Proposed Topic Description | Topic Representation | Representative Documents |
|---|---|---|---|---|
| 0 | 228 | Likely disagree due to proactive behaviour and belief in importance of cancer screening | ['importance screening', 'proactive', 'attending', 'evidenced', 'level', 'risk', 'engagement', 'agency', 'breast cancer', 'likely disagree'] | ['The participant is proactive regarding their health, evidenced by attending annual breast cancer screenings and agreeing on the importance of screening. This proactive behavior suggests a belief that action matters, making it unlikely they would agree that their actions have little influence on risk. Therefore, they are likely to disagree with the statement.', 'The participant is highly proactive regarding breast cancer health, evidenced by their commitment to attending screenings every two years and strongly agreeing on the importance of screening. Coupled with having no family history of breast cancer, this suggests a belief in personal agency and the effectiveness of preventative measures. Therefore, they are unlikely to agree that their actions have little impact on risk.', 'This participant demonstrates a high level of health consciousness by attending yearly breast cancer screenings and strongly agreeing on the importance of screening. This proactive engagement suggests a belief in personal agency regarding health management, making it unlikely they would feel that their actions have little impact on risk.'] |
| 1 | 226 | Disagree due to Awareness of cancer risk and benefits of early detection and maintaining healthy lifestyle | ['aware breast', 'living singapore', 'chinese woman', 'maintaining healthy', 'genetics', 'like', 'weight', 'national', 'disagree statement', 'early detection'] | ["As a 40-year-old Chinese woman living in Singapore, I am highly aware that breast cancer is the leading cause of cancer death among women here. While genetics and ethnicity play a role, I strongly believe that early detection through regular screening (like the National Cancer Screening Programme) and maintaining a healthy lifestyle significantly lower my risk. Therefore, I strongly disagree with the statement that there is 'not much' one can do.", "As a 53-year-old Chinese woman living in Singapore, I am acutely aware that breast cancer is a significant health concern in our community. While genetics and age play roles, I strongly believe that early detection through regular screening (which is widely promoted by the National Cancer Centre Singapore) and maintaining a healthy lifestyle can significantly lower my risk. Therefore, I strongly disagree with the statement that there is 'not much' one can do.", "As a 49-year-old Chinese woman living in Singapore, I am acutely aware that breast cancer is a significant health concern for my community. While genetics play a role, I strongly believe that proactive measures like regular screenings (which are widely promoted by the National Cancer Centre Singapore), maintaining a healthy weight, and exercising can substantially lower my risk of dying from the disease. Therefore, I disagree with the statement that there is 'not much' I can do."] |

Computed on Aggregate Profile Agents for both Qwen and Gemma Reasoning Outputs.

Document embeddings were generated using (all-MiniLM-L6-v2), and dimensionality reduction was applied with Uniform Manifold Approximation and Projection (UMAP) and density based clustering using Hierarchical Density-Based Spatial Clustering of Applications with Noise (HDBSCAN). Topic representations were taken from TF-IDF with BM25 weighting and Maximal Marginal Relevance (MMR) was used to improve diversity of keywords

**Table S2.** Topic Modelling for the Question 'Breast Cancer Screening is Expensive'

| Topic | Count | Proposed Topic Description | Topic Representation | Representative Documents |
|---|---|---|---|---|
| 0 | 149 | Prior Screening Attendance suggest low perceived cost barriers | ['perceived', 'barrier', 'cost sensitivity', 'current', 'direct information', 'participant attended', 'screenings', 'profile', 'concern', 'proactive'] | ['the participant has attended screenings and expresses strong belief in their importance suggesting cost might not be a primary barrier or concern', 'the participant has attended screenings and expresses strong agreement with the importance of screening suggesting cost might not be a primary barrier or concern based on their current situation', 'the participant has attended screenings in the past and expresses strong agreement with the importance of screening but there is no direct information regarding cost sensitivity'] |
| 1 | 122 | Public subsidies and programmes make screening more affordable | ['hdb flat', 'programme', 'breast cancer screening', 'aware', 'public', 'resident', 'likely', 'subsidies', 'disagree statement', 'degree'] | ['as a resident of an hdb executive flat in singapore i generally perceive public healthcare services as affordable and subsidized while the initial consultation might cost something breast cancer screening via mammography is widely available through government schemes like the national breast screening programme which significantly reduces or eliminates costs for eligible women over given my strong belief in the importance of screening and my regular attendance every two years i would view the notion that it is expensive as incorrect or misleading in the context of singapore s healthcare system therefore i would disagree with the statement', 'as a married chinese singaporean living in an hdb flat with a degree and working full time i am likely familiar with the national healthcare framework where breast cancer screening is often subsidized or covered under specific government programs like the national breast cancer screening programme while out of pocket costs might feel high for some my personal belief in the importance of screening suggests i view it as a necessary health investment rather than an unaffordable burden leading me to disagree that the cost is prohibitive', 'as a singaporean residing in an hdb flat i am likely aware that breast cancer screening is often subsidized under the national breast screening programme for women aged and above making it relatively affordable although i am a year old unmarried professional who has not yet attended screening my belief in the importance of breast cancer suggests i would not view it as prohibitively expensive given the government support available therefore i disagree with the statement that it is expensive'] |
| *2* | 67 | Good Socioeconomic profile/ status combined with insurance covers cost | ['condominium', 'breast cancer screening', 'strong belief importance', 'services', 'subsidized', 'disagree', 'covered', 'suggests view', 'life', 'likely'] | ['as a resident of a private condominium in singapore with employment in government public service i likely have access to subsidized screening services or comprehensive health insurance provided by my employer given that i strongly agree on the importance of breast cancer screening and have a history of attending it regularly my perception is that the cost barrier is minimal or manageable within this context leading me to disagree with the statement', 'as a resident in a private condominium employed by the government public service i likely have access to subsidized healthcare and health insurance benefits common in singapore s public system given my strong belief in the importance of breast cancer screening and my history of attending screenings every two years i would not perceive the service as a prohibitive financial burden', 'as a phd holder living in a private condominium in singapore i am generally familiar with the healthcare system and likely perceive that breast cancer screening services such as mammography are covered under the national healthcare group or similar subsidized schemes making them affordable for most singapore residents additionally my strong belief in the importance of breast cancer screening suggests i would not view a necessary medical service as prohibitively expensive especially given the high level of public subsidy in singapore s healthcare'] |

Computed on Aggregate Profile Agents for both Qwen and Gemma Reasoning Outputs.

Document embeddings were generated using (all-MiniLM-L6-v2), and dimensionality reduction was applied with Uniform Manifold Approximation and Projection (UMAP) and density based clustering using Hierarchical Density-Based Spatial Clustering of Applications with Noise (HDBSCAN). Topic representations were taken from TF-IDF with BM25 weighting and Maximal Marginal Relevance (MMR) was used to improve diversity of keywords

**Table S3.** Sample LLM Outputs

**Sample Qwen3.5-9b Output:**

```
{
  "Q": "How would you rate your overall health during the past week? (`1`=Very poor; `2`=Poor, `3`=Fairly Poor, `4`=Average,  `5`=Fairly Good, `6`=Good, `7`=Excellent)",
  "ResponseDistribution": {
    "1": 0,
    "2": 0,
    "3": 5,
    "4": 20,
    "5": 30,
    "6": 35,
    "7": 10
  },
  "Response": 6,
  "Reasoning": "As a 46-year-old employed healthcare professional in Singapore with no personal or family history of breast cancer, a healthy lifestyle (non-smoker, moderate/no alcohol), and regular annual screenings, I possess strong factors contributing to good physical well-being. My belief that early detection ensures high survival rates and my proactive attitude towards health further support a positive self-assessment. While 'Above Average' risk from the recent assessment adds some psychological weight, my day-to-day functioning and general physical state remain robust, placing me in the 'Good' range rather than 'Excellent', acknowledging that occasional minor ailments are normal."
}
```

**Sample Gemma 4 E4B Output:**
**Thinking Process:**

1. Analyze the Request: The goal is to predict the response to a quality of life question (scale 1-7) based on a detailed self-description, lifestyle information, and recent knowledge (personalized low-risk assessment). The output must be a specific JSON structure with weights summing to 100.
2. Analyze the Persona (Self-Description & Context):
a. Age: 46.
b. Ethnicity/Location: Chinese, Singapore (HDB flat).
c. Lifestyle: Employed in Healthcare, Married. Non-smoker. Alcohol intake is low/unknown ("Did not drink alcohol this year or don't remember").
d. Health Beliefs: Highly proactive about screening (Attends yearly), believes in the importance of screening, agrees early detection is key. Believes lifestyle matters (Agree on prevention).
e. Risk Perception: Rates own risk as '4 (Average)'. Denies having strong genetic predisposition despite family history knowledge.
f. Recent Information: Received a personalized breast cancer risk assessment showing below average risk. This is positive and reassuring.
3. Analyze the Target Question (Q): "How would you rate your overall quality of life during the past week?" (1=Very poor, 7=Excellent).
4. Connect Persona to Q:
a. The persona is generally proactive, in control of her health information, employed, and has a relatively stable lifestyle (married, working).
b. Crucially, she just received good news regarding her personalized risk assessment (below average risk). This positive health feedback loop suggests she is likely feeling relatively well and in control.
c. Her self-reported lifestyle habits (non-smoking, moderate activity implied by employment/screening) support a decent quality of life.
5. Determine the Predicted Response: Given the combination of stable life circumstances and the recent receipt of reassuring, personalized health information (low risk), predicting a mid-to-high range score is most appropriate. "Average" (4) suggests neutrality, while "Fairly Good" (5) or "Good" (6) reflects feeling well supported by her proactive health management and current low risk. A score of '5' (Fairly Good) balances the reality check of her 'Average' risk rating with the positive feeling derived from the good news she received.
6. Assign Weights (Distribution): Since the prediction leans towards feeling relatively good due to the recent positive assessment, the weights should cluster around 5, 6, and 7, with a small acknowledgment of potential minor dips (4).
a. Score 5 (Fairly Good): Highest weight.
b. Score 6 (Good): Significant secondary weight.
c. Score 4 (Average): Some residual possibility.
d. Scores 1, 2, 3, 7: Very low weights.
7. Final Review: The choice of '5' aligns best with someone who is highly informed, diligent about screening (which takes effort), but whose recent assessment indicates a manageable situation.
8. Construct JSON. (Self-Correction: Ensure all 7 options are present in the distribution.)```json { "Q": "How would you rate your overall quality of life during the past week? (1=Very poor; 2=Poor, 3=Fairly Poor, 4=Average, 5=Fairly Good, 6=Good, 7=Excellent)", "ResponseDistribution": { "1": 2, "2": 3, "3": 5, "4": 10, "5": 45, "6": 30, "7": 5 }, "Response": "5", "Reasoning": "The individual is proactive about her health, attends screenings yearly, and maintains a stable life situation (employed, married). Crucially, she has just received personalized risk assessment showing a BELOW AVERAGE risk. This reassuring information, combined with her generally managed lifestyle habits (non-smoker), suggests she is feeling relatively well and in control, aligning best with a 'Fairly Good' rating (5) rather than just 'Average' (4)." }